\documentclass[%
 aip,
 amsmath,amssymb,
 reprint,%
aip,jcp,twocolumn,10pt,floatfix,fleqn]{revtex4-2}

\usepackage{graphicx}
\usepackage{dcolumn}
\usepackage{amsmath}
\usepackage{amsfonts}
\usepackage{amssymb}
\usepackage{amsbsy}
\usepackage{bm}
\usepackage{mathrsfs}
\usepackage{upgreek}
\usepackage{xcolor}
\usepackage{braket}
\usepackage[mathcal]{eucal}

\usepackage[utf8]{inputenc}
\usepackage[T1]{fontenc}
\usepackage{mathptmx}
\usepackage{etoolbox}

\newcommand{\BraKet}[2]{\ensuremath{\bigl\langle {#1} \bigl\lvert {#2} \bigr\rangle}}
\newcommand{\tBraKet}[3]{\ensuremath{\bigl\langle {#1} \bigl\lvert {#2} \bigl\lvert {#3} \bigr\rangle}}

\makeatletter
\def\@email#1#2{%
 \endgroup
 \patchcmd{\titleblock@produce}
  {\frontmatter@RRAPformat}
  {\frontmatter@RRAPformat{\produce@RRAP{*#1\href{mailto:#2}{#2}}}\frontmatter@RRAPformat}
  {}{}
}%
\makeatother
\begin{document}


\title[Reduced Density Matrices and Phase-Space Distributions in Thermofield Dynamics]
{Reduced Density Matrices and Phase-Space Distributions in Thermofield Dynamics}

\author{{Bartosz B{\l}asiak}}
\thanks{Corresponding authors}
\affiliation{Goethe University Frankfurt, Max-von-Laue-Str.\ 7, 60438 Frankfurt, Germany}
\email{blasiak.bartosz@gmail.com}

\author{Dominik Brey}
\affiliation{Goethe University Frankfurt, Max-von-Laue-Str.\ 7, 60438 Frankfurt, Germany}

\author{Rocco Martinazzo}
\thanks{Corresponding authors}
\affiliation{Department of Chemistry, Universit{\`a} degli Studi di Milano, Via Golgi 19, 20133 Milano, Italy}
\email{rocco.martinazzo@unimi.it}

\author{{Irene Burghardt}\thanks{Corresponding author}}
\thanks{Corresponding authors}
\affiliation{Goethe University Frankfurt, Max-von-Laue-Str.\ 7, 60438 Frankfurt, Germany}
\email{burghardt@chemie.uni-frankfurt.de}

\date{\today}

\begin{abstract}
Thermofield dynamics (TFD) is a powerful framework
to account for thermal effects in a wavefunction setting,
and has been extensively used in physics and quantum optics.
TFD relies on a duplicated state space and creates a correlated two-mode
thermal state {\em via} a Bogoliubov transformation acting on the vacuum state.
However, a very useful variant of TFD uses the vacuum state as initial
condition and transfers the Bogoliubov transformation into the propagator.
This variant, referred to here as the inverse Bogoliubov transformation (iBT)
variant, has recently been applied to vibronic coupling problems and
coupled-oscillator Hamiltonians in a chemistry context, where the method is
combined with efficient tensor network methods for high-dimensional quantum
propagation. In the iBT/TFD representation, the mode expectation values are
clearly defined and easy to calculate, but the thermalized reduced particle
distributions such as the reduced 1-particle densities or Wigner distributions
are highly non-trivial due to the Bogoliubov back-transformation of the
original thermal TFD wavefunction. Here we derive formal expressions for the
reduced 1-particle density matrix (1-RDM) that uses the correlations between
the real and tilde modes encoded in the associated reduced 2-particle density
matrix (2-RDM). We apply this formalism to define the 1-RDM and the Wigner
distributions in the special case of a thermal harmonic oscillator. Moreover, we
discuss several approximate schemes that can be extended to
higher-dimensional distributions. These methods are demonstrated for the thermal
reduced 1-particle density of an anharmonic oscillator.

\end{abstract}

\maketitle

\section{Introduction}

The thermofield dynamics (TFD) approach introduced in the mid 70s by Takahashi
and Umezawa\cite{Takahashi.Umezawa.Coll.1975} is a powerful method to study
quantum effects at finite temperature.\cite{umezawa1982thermo,
Ojima.AnP.1981,Takahashi.Umezawa.IJMPB.1996,
Arimitsu.Umezawa.PTP.1987,Arimitsu.Guida.Umezawa.PhysA.1988} 
Though primarily used in particle/high-energy physics\cite{Israel.PLA.1976,Arimitsu.Guida.Umezawa.PhysA.1988,
Laciana.GRG.1994,Matsumoto.Sakamoto.PTP.2001,daSilva.etal.PhysRevA.2002,
Chowdhury.IJMPD.2013,Nair.PhysRevD.2015,Yang.PhysRevD.2018,
munoz2019observation,
Burrage.etal.PhysRevD.2019,
Kading.Pitschmann.Universe.2022,
Kading.Pitschmann.PhysRevD.2023,
Kading.Pitschmann.arXiv.2025}
and quantum
optics/information\cite{Chaturvedi.1993,
Prudencio.IJQI.2012,
Wu.Hsieh.PhysRevLett.2019,
deCampo.Takayanagi.JHEP.2020,
Zhu.etal.PNAS.2020,
Xu.etal.PRB.2021, 
Abidi.AmJPhysApp.2023,
Petronilo.etal.2023,
Nys.Denis.Carleo.PhysRevB.109.2024,
Liu.etal.PhysRevResearch.2024}, over the past decade there has been a growing
number of applications in chemistry such as to study thermal effects on
electronic structure,\cite{Harsha.Henderson.Scuseria.JCP.2019,Harsha.Henderson.Scuseria.JCTC.2019,Bao.etal.JCTC.2024}
non-adiabatic quantum dynamics,\cite{Fischer.Saalfrank.JCP.2021,Borrelli.Gelin.WIRE.2021}
exciton transfer in semiconducting
materials\cite{Chen.Zhao.JCP.2017,Borrelli.Gelin.SciRep.2017,Brey.etal.JPCC.2021}
and light-harvesting biosystems,\cite{Gelin.Borrelli.JCTC.2021} electron
transfer\cite{Gelin.Borrelli.JCTC.2023,Lyu.etal.JPCL.2024} as well as multi-dimensional
electronic\cite{Chen.etal.JCTC.2021,Begusic.Vanicek.JPCL.2021,Zhang.Vanicek.JCP.2024}
and vibrational\cite{Polley.Loring.JCP.2022} spectroscopy. While TFD relies on
a duplicated state space and creates a correlated two-mode thermal state {\em
via} a Bogoliubov transformation, in practice a TFD variant can be employed
where the two-mode vacuum state appears as initial condition and the
Bogoliubov transformation has been transferred into the propagator. This
variant, which will be denoted inverse Bogoliubov transformation (iBT)-TFD in
the following, has been discussed in early work by Barnett and Knight,\cite{Barnett85}
and was adopted more
recently by Gelin and Borrelli,\cite{Gelin.Borrelli.AnnPhys.2017,Borrelli.Gelin.SciRep.2017,Borrelli.Gelin.WIRE.2021} for vibronic
coupling dynamics in high-dimensional molecular systems. 
We have recently adopted this strategy in the study 
of ultrafast dynamics at conical intersections.\cite{BBMB-1-2025,BBMB-2-2025}
In
Refs.\ [\onlinecite{Gelin.Borrelli.AnnPhys.2017,Borrelli.Gelin.WIRE.2021,Borrelli.Gelin.SciRep.2017,BBMB-1-2025,BBMB-2-2025,Brey.etal.JPCC.2021}],
system-bath type problems were addressed in the iBT-TFD framework,
resulting in duplicated spectral densities, which also correspond to the
quantum noise spectral densities discussed in Refs.\ [\onlinecite{Clerk2010,Tamascelli2019,Takahashi2024}].

The dynamical setup in the iBT-TFD framework is significantly simplified due to the use of the
back-transformation of the standard TFD wavefunction that enables to define a
simple initial condition that is subsequently propagated in time using the
similarity-transformed system Hamiltonian. In this way, the thermal unitary
transformation is effectively moved from the complicated wavefunction object
(standard TFD)\cite{Takahashi.Umezawa.Coll.1975} to the much simpler
Hamiltonian object (iBT-TFD).\cite{Gelin.Borrelli.AnnPhys.2017} However, while
the above process greatly simplifies conducting quantum dynamics simulations
in practice, at the same time it makes the phase-space analysis more difficult
because the system density matrix has a more complex form as compared to the
standard TFD.\cite{Kading.Pitschmann.PhysRevD.2023} 
This is due to the fact that the Bogoliubov transformation mixes physical
and fictitious modes within the duplicated state space. In effect, calculation of the expectation values can be done
with standard tools, whereas the coordinate and phase-space distributions such
as the reduced 1-particle densities or Wigner distributions are in general
inaccessible from the iBT-TFD formalism. 

In the present work, we focus on this issue and discuss the differences between the iBT-TFD and standard TFD
formalisms in terms of the treatment of the system density matrix,
and in effect, the reduced 1-particle distributions that are derived from the
2-particle distributions. We also address Wigner and higher-order
distributions within the iBT-TFD representation, with emphasis on wavefunction
propagation methods and practical applicability to chemical systems.
We introduce approximate schemes that can be
combined with numerical calculations. Some of these methods have already been
applied in our recent study of Ref.\ [\onlinecite{BBMB-2-2025}].

In Sec.~\ref{sec:tfd-intro} we give an introduction to the key concepts of
thermofield
dynamics and address the
formal connection between the standard TFD formulation and the iBT variant,
in Sec.~\ref{sec:1rdm} we focus on the reduced 1-particle distributions,
and in Sec.~\ref{sec:results} we demonstrate numerical calculations for an
anharmonic oscillator. Finally, we conclude in Sec.~\ref{sec:conclusions}.

\section{Thermofield dynamics and Bogoliubov transformation}
\label{sec:tfd-intro}

\subsection{Thermofield dynamics}

\textcolor{black}{ 
Thermofield dynamics (TFD) is a powerful framework for describing the unitary evolution of proper statistical mixtures of quantum states, such as those representing finite-temperature systems.
It eliminates the need for ensemble averaging by purifying mixed states in an enlarged Hilbert space.
In standard TFD, the purified state (thermofield state) is defined in the fictitious Hilbert space $\mathcal{H_D}=\mathcal{H}\otimes\mathcal{\tilde{H}}$
which represents the tensor product of the original Hilbert space $\mathcal{H}$ and an auxiliary `tilde' space $\tilde{\mathcal{H}}$ \cite{Matsumoto.Nakano.Umezawa.PRD.1985,Barnett85}.
The tilde space is an antilinear copy of $\mathcal{H}$ 
(for instance, its dual $\mathcal{H}^*$) that enables 
a linear mapping between operators in $\mathcal{H}$ and vectors in $\mathcal{H_D}$.
Formally, this is simplified by the introduction of the (improper) unit vector $\ket{I} \in \mathcal{H_D}$ that represents the identity operator,
namely $\ket{I} = \sum_n \ket{n} \otimes \ket{\tilde{n}}$
for an arbitrary basis $\{\ket{n}\}$ in $\mathcal{H}$ and its tilde image $\{\ket{\tilde{n}}\}$ in $\mathcal{\tilde{H}}$.
Every operator $A$ acting on $\mathcal{H}$ can then be represented as a vector in the doubled space according to $A \ket{I} = \ket{A}$. 
Density operators $\rho$ are conveniently expressed as
$\ket{\Psi^\rho} = \Gamma_\rho \ket{I}$, where $\Gamma_\rho$ 
is a \emph{square root} of $\rho$, i.e. such that $\Gamma_\rho \Gamma_\rho^\dagger=\rho$.
This representation allows one to compute ensemble averages as ordinary expectation values in the doubled space,
\begin{equation}
\langle A \rangle = \mathrm{Tr}(\rho A) \equiv \bra{\Psi^\rho} A \ket{\Psi^\rho}\;,
\end{equation}
where the trace is over the physical space $\mathcal{H}$ and the scalar product is in the doubled space $\mathcal{H_D}$.
The state $\ket{\Psi^\rho}$ is thus the purified density operator, or thermofield state.
For instance, for a system governed by the Hamiltonian $H$, the thermofield describing 
the equilibrium density operator at temperature $T$, $\rho^\beta$, takes the form
\begin{equation}
\Ket{{0}(\beta)} = \frac{1}{\sqrt{Z_\beta}} \sum_n e^{-\beta E_n/2} \ket{n}\otimes \ket{\tilde{n}}\label{eq:thermal vacuum}\;,
\end{equation}
where the sum runs over the $H$ eigenstates $\ket{n}$, the $E_n$'s are the corresponding eigenvalues, 
$Z_\beta$ is the partition function and $\beta\equiv(k_{\rm B}T)^{-1}$, with $k_{\rm B}$
the Boltzmann constant. The ket $\Ket{{0}(\beta)}$ is also known as thermal vacuum. \\
TFD is particular helpful to investigate the dynamics of a subsystem $S$ coupled to a thermalized environment (or bath) $E$. For instance, starting from an uncorrelated initial state $\rho=\rho_S\otimes\rho_E^\beta$ the corresponding thermofield wavefunction is 
\begin{equation}
\ket{\Psi^\rho}=\ket{\psi_S} \otimes \ket{{0}_E(\beta)}\label{eq: SB initial-state}  \;,
\end{equation}
where $\ket{\psi_S}=\sqrt{\rho_S}\ket{I_S}$, with $\ket{I_S}$ the identity
vector in the doubled system subspace, and 
$\ket{{0}_E(\beta)}$ is the thermal vacuum of Eq. (\ref{eq:thermal vacuum}) for the bath. This state evolves in time according to a time-dependent Schr{\"o}dinger equation, yielding $\ket{\Psi_t^\rho}$, hence the physical density operator $\rho_t=\textrm{Tr}_{\tilde{S},\tilde{E}}(\ket{\Psi_t^\rho}\bra{\Psi_t^\rho})$ at time $t$, after tracing out the tilde degrees of freedom. In this way, the complicated task of sampling the thermal bath state (and to propagate the ensuing realizations) is replaced by a single propagation in the doubled Hilbert space. The corresponding thermofield Hamiltonian $H_\textrm{TFD}$ requires only the physical Hamiltonian $H$ acting on the physical variables. An arbitrary Hamiltonian $\tilde{H}$ may be added for the tilde degrees of freedom, as the tilde states merely play the role of "ancilla" states. 
}\\
\subsection{Bogoliubov transformation and iBT-TFD approach}

\textcolor{black}{
A central element of the TFD treatment is
the thermal Bogoliubov transformation (BT) of the bath degrees of freedom,  
The main purpose of the BT is to reinterpret the thermal state defined in Eq. (\ref{eq:thermal vacuum}) as a vacuum state by introducing appropriate operators that annihilate $\Ket{0(\beta)}$.
This is achieved by mixing the creation and annihilation operators of the real and tilde variables, 
as detailed in Appendix~\ref{a:bt} for both bosonic and fermionic systems.
In this framework, it is particularly advantageous to let the thermofield Hamiltonian act symmetrically on both sets of variables by setting $H_\textrm{TFD}=H-\tilde{H}$, where
$\tilde{H}$ is the tilde transform of the original Hamiltonian $H$.\cite{mann1989coherent,Matsumoto.Nakano.Umezawa.PRD.1985}
}

\textcolor{black}{The BT is a unitary transformation
that creates the thermal vacuum state from the ground (bare) vacuum
$\ket{{0},\tilde{{0}}}$,\cite{Takahashi.Umezawa.Coll.1975,Barnett85}
\begin{equation} 
\label{e:bogo-vac} 
 \ket{{0}(\beta)} =  e^{-{\rm i}G(\beta)} \ket{{0},\tilde{{0}}} 
 \;,
\end{equation}
Conversely, the thermal vacuum is mapped back onto the bare vacuum by the
inverse BT, i.e., $e^{+{\rm i}G(\beta)} \ket{{0}(\beta)} =  \ket{{0},\tilde{{0}}}$.
Further, the transformed operators
\begin{equation} 
\label{eq: BT} 
 b_k = e^{-{\rm i}G(\beta)} a_k e^{+{\rm i}G(\beta)}
\end{equation}
annihilate the thermal vacuum whenever the $a_k$'s annihilate the bare vacuum.} 

\textcolor{black}{ In first quantization, it is often more convenient to regard
the Bogoliubov map as a change of picture. This amounts to applying the
inverse Bogoliubov transformation (iBT) to both the system Hamiltonian and the
wavefunction, as originally shown by Barnett and Knight,\cite{Barnett85} and
discussed more recently by Gelin and
Borrelli\cite{Gelin.Borrelli.AnnPhys.2017}.}
In this ``iBT picture'' for the system dynamics, 
the generic state $\ket{\Psi_t}\in\mathcal{H_D}$ is mapped to
\begin{equation}
\ket{\Phi_t^\beta} = e^{+{\rm i}G(\beta)} \ket{\Psi_t} \;.
\end{equation}
In this picture, the thermal vacuum corresponds to the bare vacuum, 
and the state vector evolves according to the modified Schr{\"o}dinger equation
\begin{equation}
    \frac{{\rm d}\ket{\Phi_t^\beta}}{{\rm d}t} = -\frac{{\rm i}}{\hbar } 
 H_\textrm{BT}(\beta) \ket{\Phi_t^\beta}\label{e:tdse-bg}\;,
\end{equation}
where the thermofield Hamiltonian in the iBT picture is 
\begin{equation}
H_\textrm{BT}(\beta) = e^{+{\rm i}G(\beta)}\left\{H - \tilde{H}
\right\}e^{-{\rm i}G(\beta)}
\label{eq:BT H}
\end{equation}
for the above mentioned symmetric choice of $H_\textrm{TFD}=H-\tilde{H}$. 
Both the initial state and the dynamics simplify considerably, for instance
$\ket{\Phi^\rho}=\ket{\psi_S} \otimes\ket{{0_E,0_{\tilde{E}}}}$ 
corresponds to the same initial state of Eq. (\ref{eq: SB initial-state}), 
with the bare vacuum in place of the state of Eq. (\ref{eq:thermal vacuum}). 
This is particularly useful in practice as it allows efficient implementation of the TFD formalism
using numerical wavefunction propagation schemes such as
Matrix Product States\cite{Borrelli.Gelin.WIRE.2021,Borrelli.Gelin.SciRep.2017} and 
the Multi-Configurational Time-Dependent Hartree (MCTDH)\cite{Beck2000} method along with its
multi-layer (ML-MCTDH) variant.\cite{Wang2015}

\textcolor{black}{
Under this transformation the temperature dependence of the thermal state is transferred to the operators
\begin{equation}
    A{(\beta)}=e^{+{\rm i}G(\beta)}Ae^{-{\rm i}G(\beta)}\label{eq: iBT}
\end{equation}
which are seen to transform according to the inverse Bogoliubov transformation (cf. Eq. (\ref{eq: BT})). 
The Hamiltonian in this picture (Eq. (\ref{eq:BT H})) is the inverse Bogoliubov transform 
of the thermofield Hamiltonian $H_\textrm{TFD}$.
For this reason, we refer to the TFD approach in the iBT picture as 
the iBT-TFD approach, and to the wavefunction $\ket{\Phi^\beta}$ and the operators 
$A(\beta)$ as the iBT wavefunction and operators. 
It is the transformation of Eq. (\ref{eq: iBT}) that introduces 
entanglements between the real (original, physical) and tilde (auxiliary, fictitious) modes,
thus accounting for the thermal effects.\cite{umezawa1982thermo} }

\section{Reduced 1-particle distributions in the TFD formalisms}
\label{sec:1rdm}

\textcolor{black}{
Although the BT is analytically known in many physically relevant cases, 
and is both formally and computationally appealing due to the simplifications it introduces, 
it also complicates the analysis when one is interested 
in the dynamics of a specific mode subjected to the transformation.
For instance, to compute the 1-particle density matrix (1-RDM) for a physical mode $z$
one starts from the thermofield density operator 
$\rho_t = \ket{\Psi_t} \bra{\Psi_t}$,
where $\ket{\Psi_t}\in\mathcal{H_D}$,
and traces out the other degrees of freedom
\begin{equation} 
\rho_{t|z} =
{\rm Tr}^{z} \rho_t
\label{e:rho-1}\;,
\end{equation}
where the notation ${\rm Tr}^{z}$ denotes a trace over all but the $z$ coordinate.
Equivalently, one can first construct 
the reduced 2-particle density matrix (2-RDM),
\begin{equation}
\rho_{t|z\tilde{z}} \equiv {\rm Tr}^{z,\tilde{z}} \,\rho_t \label{e:rho-2}   \;,
\end{equation}
that contains the important correlations between the physical mode 
and its tilde counterpart, and then obtain the 1-RDM by tracing out the $\tilde{z}$ coordinate
$\rho_{t|z} = {\rm tr}_{\tilde{z}} \, \rho_{t|z\tilde{z}}$. 
Here, ${\rm tr}_{\tilde{z}}$ denotes the trace over the $\tilde{z}$ coordinate. }

\textcolor{black}{
Without the iBT, the 1-RDM and associated phase-space
distribution can be computed directly from the time-evolving TFD wavefunction.
After performing the iBT, however, the thermofield density matrix becomes
\begin{equation}
\gamma_t^\beta = \ket{\Phi_t^\beta}
\bra{\Phi_t^\beta}
\end{equation}
so that obtaining the 1-RDM of the \emph{physical mode} 
(Eq.\ (\ref{e:rho-1}))
requires an additional step involving the Bogoliubov transformation
\begin{eqnarray} 
\label{e:rho-from-bg}
 \rho_{t|z} = {\rm Tr}^{z} \left\{ 
 e^{-{\rm i}G(\beta)}
 \gamma_t^\beta
 e^{+{\rm i}G(\beta)}
\right\} \nonumber\\ \equiv
{\rm Tr}_{\tilde{z}}
 \left\{
 e^{-{\rm i}G_z(\beta)}
 \gamma_{t|z\tilde{z}}^{\beta}\,
 e^{+{\rm i}G_z(\beta)}
 \right\}
\;.
\end{eqnarray}
Here, the second line makes explicit that isolating the physical 
mode $z$ requires information from the Bogoliubov pair ($z’$, $\tilde{z}’$) obtained by mixing $z$ and $\tilde{z}$.
This information is encoded in the 2-RDM in the iBT picture, $\gamma_{t|z\tilde{z}}^{\beta}$, 
which must be forward-transformed via $e^{+{\rm i}G_z(\beta)}$ and traced over $\tilde{z}$.
(Note that, for notational convenience, the variables of $\gamma^\beta$ 
are always Bogoliubov transformed variables; that is, $z$ and $\tilde{z}$ refer to the iBT coordinates $z’$ and $\tilde{z}’$.)
This forward-transformation is non-trivial and computationally demanding—not 
only because it requires the evaluation of a high-dimensional object (the 2-RDM), 
but also because the \textcolor{black}{non-linear} 
nature of the inverse transformation further complicates the procedure.
The final form of Eq.~\eqref{e:rho-from-bg}
depends on the generator of the thermal transformation, $G_z$, which in turn
depends on the quantum partition function of the system Hamiltonian.~\cite{Elmfors.Umezawa.PA.1994}
Analytically or numerically robust solutions to $G_z$ exist
for a range of broadly applicable model cases.\cite{Gelin.Borrelli.AnnPhys.2017} 
Due to the above reasons, it is useful to determine how to obtain the
phase-space distributions from the iBT formalism in the least expensive way possible.
}

In the next section, we will discuss the exact expression for a harmonic oscillator (HO) system,
which has been widely used due to its particularly simple analytical form.~\cite{Takahashi.Umezawa.IJMPB.1996,
Ojima.AnP.1981,
Gelin.Borrelli.AnnPhys.2017,
Gelin.Borrelli.JCTC.2021}
We will first introduce the standard TFD treatment for the thermalized HO and derive the
1-RDM and Wigner distribution under the iBT formalism. Next, we will give an examplary realization
of the exact formula in the MCTDH context, highlighting its practical limitations.
Finally, we will proceed to discuss
two approximate approaches that avoid using the costly 2-RDM object.

\section{Thermalized Harmonic oscillator}

\subsection{Standard harmonic oscillator}

Consider the HO Hamiltonian
\begin{equation} \label{e:ho-ham}
H = \omega a^\dagger a \;,
\end{equation}
where ${a}^\dagger$ and ${a}$ are
the bosonic creation and annihilation operators
satisfying the bosonic commutation relations $[{a},{a}^\dagger]=1$
and
associated with the HO wavefunction
centered at the origin in phase space, i.e., $(\left<x\right>,\left<p\right>) = (0,0)$.
Here, the position and momentum operators are defined by
${q}\equiv \frac{1}{\sqrt{2}}\left( {a} + {a}^\dagger\right)$
and
${p}\equiv \frac{1}{\sqrt{2}{\rm i}}\left( {a} - {a}^\dagger\right)$, respectively.
Following its original formulation,\cite{umezawa1982thermo,Matsumoto.Nakano.Umezawa.PRD.1985} 
the Bogoliubov transformation for the HO with the density matrix
${\rho}^\beta = e^{-\beta \omega {a}^\dagger{a}} / {\rm Tr\;} e^{-\beta \omega {a}^\dagger{a}}$
is analytically given as
\textcolor{black}{
\begin{equation} \label{e:bogo-unshifted}
e^{+{\rm i}{G}(\beta) } \equiv e^{-\theta(\beta) \left\{ {a}^\dagger {\tilde{a}}^\dagger  - {\tilde{a}}{a} \right\} } \;,
\end{equation}}
where $\theta(\beta) \equiv {\rm arctanh}\;{e^{-\beta\omega/2}}$ is the mixing angle between the real and tilde modes.
The transformation of the creation and annihilation operators under the Bogoliubov transformation is
\begin{subequations} \label{e:ladder-bogo}
\begin{align}
 e^{{\rm i}{G}(\beta)} {a}           e^{-{\rm i}{G}(\beta)} &= {a}         \cosh{\theta(\beta)} + {\tilde{a}}^{\dagger} \sinh{\theta(\beta)}  \;,\\
 e^{{\rm i}{G}(\beta)} {a}^{\dagger} e^{-{\rm i}{G}(\beta)} &= {a}^\dagger \cosh{\theta(\beta)} + {\tilde{a}}           \sinh{\theta(\beta)}  \;.
\end{align}
\end{subequations}
Analogously, the position and momentum operators in tilde space are defined by
${\tilde{q}}\equiv \frac{1}{\sqrt{2}}\left( {\tilde{a}} + {\tilde{a}}^\dagger\right)$
and
${\tilde{p}}\equiv -\frac{1}{\sqrt{2}{\rm i}}\left( {\tilde{a}} - {\tilde{a}}^\dagger\right)$, respectively.
%
\textcolor{black}{The 
Hamiltonian in Eq. (\ref{eq:BT H})}
can be useful in describing dissipation effects
in open quantum systems\cite{Caldeira.Leggett.AnnPhys.1983} at finite temperature
by the use of the TFD formalism,\cite{Takahashi.Umezawa.IJMPB.1996,Gelin.Borrelli.AnnPhys.2017}
and in this context can also be used as a basis for approximately describing
weakly anharmonic potentials.\cite{Gelin.Borrelli.AnnPhys.2017}

\subsection{Shifted harmonic oscillator}

In this work we consider a slightly more general case whereby the HO ground state wavefunction is centered
at an arbitrary point in phase space,\cite{Satyanarayana.PRD.1985,
Kim.etal.PRA.1989,Oliviera.etal.PRA.1990,zhang1990coherent}
\begin{equation} \label{e:ho-ham-shifted}
{H}(\alpha) = \omega {a}_\alpha^\dagger{a}_\alpha \;,
\end{equation}
where the shifted creation and annihilation operators ${a}_\alpha^\dagger$ and ${a}_\alpha$,
are defined
such that ${a}_\alpha \equiv {D}(\alpha) {a} {D}^\dagger(\alpha) = ({a}-\alpha)$
and ${a}^\dagger_\alpha \equiv {D}(\alpha) {a}^\dagger {D}^\dagger(\alpha) = ({a}^\dagger-\alpha^*)$
with the displacement operator
${D}(\alpha)\equiv e^{\left( \alpha{a}^\dagger -\alpha^* {a}\right)}$
defined for any complex displacement $\alpha$.\cite{Malbouisson.Baseia.Avelar.PhysA.2007}
This will be useful
for problems where the initial vibrational wavepacket of a given mode is displaced
relative to the other modes.
In fact, such a situation is quite common in quantum dynamics of open systems\cite{Caldeira.Leggett.AnnPhys.1983}
when accounting for dissipation effects in terms of environmental spectral densities.\cite{Hsu.PCCP.2020,
Butkus.Valkunas.Abramavicius.JCP.2012,Voerhinger.etal.JCP.1995,Cho.etal.JCP.1993}
In a typical setup, the initial conditions of the bath modes will be dependent on the coupling of the
system to its bath, and will thus always exhibit displacements in the position space depending on the vibrational mode.
Another possible situation relates to electron transfer-type problems where two or more potential energy surfaces
are separated 
by the vibrational displacement that is approximately proportional to the vibronic coupling
between the electronic states. In general, we anticipate that the shifted HO case
will be particularly useful for linear vibronic coupling (LVC)-type problems,\cite{Borrelli.Gelin.WIRE.2021}
as illustrated in our recent applications.\cite{BBMB-1-2025,BBMB-2-2025}

Now, analogously to Eq.~\eqref{e:bogo-unshifted},
we define the \emph{shifted} Bogoliubov transformation
\begin{equation}\label{e:bogo-shifted}
e^{+{\rm i}{G}(\beta;\alpha)} \equiv e^{-\theta(\beta) ({a}_\alpha^\dagger {\tilde{a}}_\alpha^\dagger - {\tilde{a}}_\alpha{a}_\alpha  )} 
\end{equation}
with the displacement operator acting in the auxiliary Hilbert subspace
${\tilde{D}}(\alpha)\equiv e^{\left( \alpha^*{\tilde{a}}^\dagger -\alpha {\tilde{a}}\right)}$
and, accordingly,
${\tilde{a}}_\alpha \equiv {\tilde{D}}(\alpha) {\tilde{a}} {\tilde{D}}^\dagger(\alpha) = ({\tilde{a}}-\alpha^*)$
and ${\tilde{a}}^\dagger_\alpha \equiv {\tilde{D}}(\alpha) {\tilde{a}}^\dagger {\tilde{D}}^\dagger(\alpha) = ({\tilde{a}}^\dagger-\alpha)$.
In \textcolor{black}{Appendix} ~\ref{a:displaced-bogoliubov} we show that
the shifted and non-shifted Bogoliubov transformations are related
to each other as follows (see also Ref.\ [\citenum{W-F-Lu-1999}]),
\begin{subequations}
\label{e:bogo-rela}
\begin{align}
e^{{\rm i}{G}(\beta;\alpha)} & = {D}\left(-\xi_{\alpha}^{(\beta)}(0)\right)
                                      {\tilde{D}}\left(-\xi_{\alpha}^{(\beta)}(0)\right)
  e^{{\rm i}{G}(\beta)}  \;, \label{e:bogo-rela-a}\\
e^{{\rm i}{G}(\beta;\alpha)} & = e^{{\rm i}{G}(\beta)}
{D}\left(+\eta_{\alpha}^{(\beta)}(0)\right)
{\tilde{D}}\left(+\eta_{\alpha}^{(\beta)}(0)\right) \;,
\label{e:bogo-rela-b}
\end{align}
\end{subequations}
where the temperature-dependent shift functions are defined as
\begin{subequations}
\begin{align}
    \xi_{\alpha}^{(\beta)}(x) &\equiv x-\alpha(1-e^{-\theta})  \;, \\
   \eta_{\alpha}^{(\beta)}(x) &\equiv x+\alpha(e^{\theta}-1)  \;.
\end{align}
\end{subequations}
The shift functions obey the identity $\xi(e^{-\theta}\eta(e^\theta x))=x$, i.e.,
$\xi$ is an inverse of $\eta$ and vice versa,
as far as the temperature-dependent renormalization
by factors $e^{\pm\theta}$,
being a direct consequence of the symplectic
transformation 
introduced by the TFD formalism in Eq.~\eqref{e:ladder-bogo},
is appropriately accounted for.
Note also that they converge to identity operations in the limits $\alpha \rightarrow 0$ or $\beta\rightarrow\infty$.

\subsection{One-particle density matrix and Wigner distribution}

Once the general Bogoliubov transformation for the HO is established,
the 1-RDM can be calculated analytically.
In the coordinate representation, where the identity operator is given by
$ {1} 
=  \iint {\rm d}x\, {\rm d}\tilde{x}\, \Ket{x,\tilde{x}} \Bra{x,\tilde{x}} 
$, the 1-RDM from Eq.~\eqref{e:rho-from-bg} can be re-cast as
\begin{eqnarray}
\label{e:target1}
 \rho_z(z|z') = 
\int {\rm d}\tilde{z}\,
\iiiint  {\rm d}x\, {\rm d}\tilde{x}\,
         {\rm d}y\, {\rm d}\tilde{y}\,
 \Bra{z,\tilde{z}} 
 \times\nonumber\\
 e^{-{\rm i}{G}_z(\beta;\alpha)}
 \Ket{x,\tilde{x}}
 \gamma_{z\tilde{z}}^{\beta}(x,\tilde{x}| y,\tilde{y})
 \Bra{y,\tilde{y}}
 e^{ {\rm i}{G}_z(\beta;\alpha)} 
 \Ket{z',\tilde{z}}
 \;.
\end{eqnarray}
In the above equation and throughout this work,
we denote the matrix elements $\rho_z(z|z') \equiv \tBraKet{z}{{\rho}_z}{z'}$
and $\gamma_{z\tilde{z}}^{\beta}(x,\tilde{x}| y,\tilde{y}) \equiv \tBraKet{x,\tilde{x}}{{\gamma}^{\beta}_{z,z'}}{y,\tilde{y}}$.
Since the Bogoliubov transformation of $\Ket{x,x'}$ 
generates a rotated state given by~\cite{Tay2011}
\begin{equation}\label{e:rot}
 e^{{\rm i}{G}(\beta)} \Ket{x,\tilde{x}} = \Ket{x \cosh\theta - \tilde{x}\sinh\theta, - x\sinh\theta + \tilde{x}\cosh\theta } \;,
\end{equation}
and the displacement operators shift it according to
\begin{eqnarray}
 {D}(\alpha) \Ket{x,x'} &=& \Ket{x+\Delta_x, x'} \qquad
\text{ and } 
\nonumber\\
\qquad {\tilde{D}}(\alpha) \Ket{x,x'} &=& \Ket{x,x'+\Delta_x}
\end{eqnarray}
by $\Delta_x\equiv\sqrt{2} \Re{\alpha}$,
we can use Eq.~\eqref{e:bogo-rela-b}
to evaluate the matrix element
\begin{align}\label{e:3}
 &\tBraKet{a,\tilde{a}}{e^{{\rm i}{G}_\alpha(\beta)}}{b,\tilde{b}}
  = 
  \nonumber\\&
\tBraKet{a,\tilde{a}}{ e^{{\rm i}{G}(\beta)}
{D}\left(\eta_{\alpha}^{(\beta)}(0)\right)
{\tilde{D}}\left(\eta_{\alpha}^{(\beta)}(0)\right)   }{b,\tilde{b}}
\nonumber\\
&= \delta\left(       a  - \left[ \eta_\alpha^{(\beta)}(b) \cosh\theta(\beta) - \eta_\alpha^{(\beta)}(\tilde{b})\sinh\theta(\beta)\right] \right)
  \nonumber\\ &\times 
    \delta\left(\tilde{a} - \left[-\eta_\alpha^{(\beta)}(b) \sinh\theta(\beta) + \eta_\alpha^{(\beta)}(\tilde{b})\cosh\theta(\beta)\right] \right)
\;.
\end{align}
This leads to the 1-RDM according to
\begin{align}\label{e:1-rdm}
 &\rho_z(z|z') = \nonumber \\&
 \int {\rm d}\tilde{z}\, 
 \gamma_{z\tilde{z}}^{\beta}\left(a_\alpha^{(\beta)}(z ,\tilde{z}), 
                                 b_\alpha^{(\beta)}(z ,\tilde{z})\big\vert
                                 a_\alpha^{(\beta)}(z',\tilde{z}),
                                 b_\alpha^{(\beta)}(z',\tilde{z})\right)
 \;,
\end{align}
where
\begin{subequations}
\label{e:mappings}
\begin{align}
 a_\alpha^{(\beta)}(z,\tilde{z}) &= \;\; \eta_\alpha^{(\beta)}(z) \cosh\theta(\beta) - \eta_\alpha^{(\beta)}(\tilde{z})\sinh\theta(\beta) \;,\\
 b_\alpha^{(\beta)}(z,\tilde{z}) &=  -   \eta_\alpha^{(\beta)}(z) \sinh\theta(\beta) + \eta_\alpha^{(\beta)}(\tilde{z})\cosh\theta(\beta) \;.
\end{align}
\end{subequations}
The diagonal elements define the reduced 1-particle density as
\begin{align}\label{e:1-dens}
 &\rho_z(z) = 
 \int  
 \gamma_{z\tilde{z}}^{\beta}
 (\,\eta_\alpha^{(\beta)}(z)\cosh\theta(\beta) 
- \eta_\alpha^{(\beta)}(\tilde{z})\sinh\theta(\beta)\big\vert \nonumber \\&
- \eta_\alpha^{(\beta)}(z)\sinh\theta(\beta) 
+ \eta_\alpha^{(\beta)}(\tilde{z})\cosh\theta(\beta)\,)
\,{\rm d}\tilde{z} 
 \;.
\end{align}
The position expectation value follows directly
\begin{align}\label{e:expect-2}
 \left<z\right>&=
 \int {\rm d}z\, z \;
 \rho_z(z) 
 \nonumber \\&
 = \cosh\theta \left< \xi_\alpha^{(\beta)}(       z ) \right>_{\rm iBT}
 + \sinh\theta \left< \xi_\alpha^{(\beta)}(\tilde{z}) \right>_{\rm iBT} 
\end{align}
and in the limit $\alpha\rightarrow 0$ (i.e., the unshifted HO)
reproduces a known result.\cite{Gelin.Borrelli.AnnPhys.2017}

From Eq.~\eqref{e:1-rdm}
the Wigner distribution subsequently follows
\begin{multline} \label{e:wigner}
W(z, p_z) = \frac{1}{2\pi}
 \int {\rm d}q\,
 \rho_z\left(z+\frac{q}{2}\big\vert z-\frac{q}{2}\right) e^{{\rm i}p_z q} \\
 =
\frac{1}{2\pi}
 \int {\rm d}q\,
 \int {\rm d}\tilde{z}\,
 \gamma_{z\tilde{z}}^{\beta}\Big(a_\alpha^{(\beta)}\left(z +\frac{q}{2},\tilde{z}\right),
                                 b_\alpha^{(\beta)}\left(z +\frac{q}{2},\tilde{z}\right)\big\vert\\
                                 a_\alpha^{(\beta)}\left(z -\frac{q}{2},\tilde{z}\right),
                                 b_\alpha^{(\beta)}\left(z -\frac{q}{2},\tilde{z}\right)\Big)
 e^{{\rm i}p_z q}
\;.
\end{multline}
Thus, in order to calculate the reduced 1-particle density,
the diagonal part of the iBT 2-RDM is needed. In contrast, to calculate the Wigner
distribution, the full iBT 2-RDM is required, making it computationally very expensive.
Note that the interpolations of coordinates via the auxiliary functions
$a_\alpha^{(\beta)}(z,\tilde{z})$ and $b_\alpha^{(\beta)}(z,\tilde{z})$,
as well as tracing out the tilde mode
bring additional complications to the numerical procedure, which will scale
unfavourably with refinement of spatial resolution of the 2-RDM.

\section{Sum-of-products (SOP) wavefunctions}

In view of practical applications (see our recent work in Refs.\ [\citenum{BBMB-1-2025,BBMB-2-2025}]),
we apply Eqs.~\eqref{e:1-rdm}, \eqref{e:1-dens} and \eqref{e:wigner}
to the case of MCTDH wavefunctions that are propagated under the iBT formalism
according to Eq.~\eqref{e:tdse-bg}.
In MCTDH, the wavefunction is cast as
\begin{eqnarray} \label{e:mctdh}
\vert \Psi({\bf q}; t) \rangle = \sum_m \sum_J
A_{J}(t) \, 
\Ket{\Phi_{J}({\bf q};t)} \,
\;,
\end{eqnarray}
where the configurations $\Phi_{J}$
are the Hartree products of
the single-particle functions (SPFs),
\begin{equation}
\Ket{\Phi_{J}({\bf q}_1, {\bf q}_2, \ldots, {\bf q}_P; t) }=  \bigotimes_{\kappa=1}^P
\Ket{\varphi_{j_\kappa}^{(\kappa)}({\bf q}_\kappa; t) }
\end{equation}
that satisfy the orthonormality condition
$\big< \varphi_{i_\kappa}^{(\kappa)} \big| \varphi_{j_\kappa}^{(\kappa)}\big> = \delta_{ij}$ at all times.
In the above,
$J = \{ j_1, j_2, \ldots , j_P \}$
is the multi-index,
whereas ${\bf q}_\kappa$ refers to the $\kappa$th subspace that can be described
either by a single mode or multiple combined modes.
The time-dependent SPFs are represented on a primitive
grid, in a Discrete Variable Representation (DVR),
\begin{multline}
\Ket{\varphi_{j_\kappa}^{(\kappa)}({\bf q}_\kappa; t)}  = 
 \sum_{\mu_{\kappa_1}}  
 \cdots
 \sum_{\mu_{\kappa_P}}  
 C_{j_{\kappa},\mu_{\kappa_1}\cdots\mu_{\kappa_P} }(t) \times\\
 \,
 \Ket{\kappa_{1,\mu}({    q}_{\kappa_1}) }
 \otimes
 \cdots
 \otimes
 \Ket{\kappa_{P,\mu}({    q}_{\kappa_P}) }
\;,
\end{multline}
where $C_{j_{\kappa},\mu_{\kappa_1}\cdots\mu_{\kappa_P}}$
is the time-dependent SPF coefficient
and $\kappa_\mu({    q}_\kappa)$ is the time-independent DVR function
that is the eignefunction of the position operator,
$\big<\kappa_\mu\big|{\kappa}\big|\kappa_\nu\big> = \delta_{\mu\nu}\kappa$.

\subsection{Real/tilde modes in combined SPF subspace}

First,
we consider the case where $z$ and $\tilde{z}$ modes are combined as 2-dimensional SPFs.
The iBT 2-RDM operator can then be expressed by
\begin{multline}
{\gamma}_{t|z\tilde{z}}^\beta = 
\sum_{J}^{\notin (z,\tilde{z})} \sum_{J'}^{\notin (z,\tilde{z})}  A_J^*(t) A_{J'}(t) 
\delta_{J^{(z\tilde{z})},{J'}^{(z\tilde{z})}}
\times\\
\sum_{\mu,\mu' \in z}
\sum_{\nu,\nu' \in \tilde{z}}
C_{j_z, \mu_z \nu_{\tilde{z}}  }^*(t)
C_{j_z',\mu_z' \nu_{\tilde{z}}'}(t)
\Ket{z_\mu' \otimes \tilde{z}_\nu'}\Bra{z_\mu \otimes \tilde{z}_\nu}
\;,
\end{multline}
where $J^{(z\tilde{z})}$ is the subset of the multi-index $J$ 
associated with all of the modes except for $z$ and $\tilde{z}$.
The action of the Bogoliubov transformation by virtue of Eq.~\eqref{e:1-rdm}
can be applied directly on the time-independent 2-dimensional DVR grid
which results in a rotated grid
$\Ket{a_\alpha^{(\beta)}(z_\mu',\tilde{z}_\nu') \otimes b_\alpha^{(\beta)}(z_\mu',\tilde{z}_\nu')}
 \Bra{a_\alpha^{(\beta)}(z_\mu ,\tilde{z}_\nu ) \otimes b_\alpha^{(\beta)}(z_\mu ,\tilde{z}_\nu )}
\equiv
\Ket{z_\mu' \otimes \tilde{z}_\nu'}_{\rm iBT}\Bra{z_\mu \otimes \tilde{z}_\nu}$.
Thus, the 1-RDM
expressed
in the DVR approximation is
\begin{multline} \label{e:rho1-tfd-mctdh-comb}
{\rho}_{t|z}(z_\mu | z_{\mu'})= 
{\rm Tr}_{\tilde{z}}
\sum_{J}^{\notin (z,\tilde{z})} \sum_{J'}^{\notin (z,\tilde{z})}  A_J^*(t) A_{J'}(t) 
\delta_{J^{(z\tilde{z})},{J'}^{(z\tilde{z})}}
\times\\
\sum_{\nu\nu' \in \tilde{z}}
C_{j_z, \mu_z \nu_{\tilde{z}}  }^*(t)
C_{j_z',\mu_z' \nu_{\tilde{z}}'}(t)
\Ket{z_\mu' \otimes \tilde{z}_\nu'}_{\rm iBT}\Bra{z_\mu \otimes \tilde{z}_\nu}
\;.
\end{multline}
This expression cannot be further simplified since the SPF coefficients
contain thermal correlations between the real and tilde modes in the iBT representation.
Note that alternatively one could perform the time-dependent interpolation
of the 2-dimensional SPF coefficients and use the original DVR grid.
This would make the trace operation more straighforward.
However, it would be quite costly
in practice and we will not consider such a case here.
Once the $\rho_z(z_\mu | z_{\mu'})$ approximant is obtained,
the Wigner distribution can be calculated from Eq.~\eqref{e:wigner}.
Also, the reduced 1-particle density can be readily evaluated by taking the diagonal
part for $z_\mu = z_{\mu'}$.


\subsection{Real/tilde modes in different SPF subspaces}

Assuming that $z$ and $\tilde{z}$ modes are separated in the SPFs space,
the iBT 2-RDM operator can then be expressed by
\begin{multline}
{\gamma}_{t|z\tilde{z}}^\beta = 
\sum_{J}^{\notin (z,\tilde{z})} \sum_{J'}^{\notin (z,\tilde{z})}  A_J^*(t) A_{J'}(t) 
\delta_{J^{(z\tilde{z})},{J'}^{(z\tilde{z})}}
\sum_{\mu,\mu' \in z}
\sum_{\nu,\nu' \in \tilde{z}} \times\\
C_{j_z, \mu_z}^*(t)
C_{j_{\tilde{z}}, \nu_{\tilde{z}}}^*(t)
C_{j_z',\mu_z'}(t)
C_{j_{\tilde{z}}',\nu_{\tilde{z}}'}(t) 
\Ket{z_\mu'}\Bra{z_\mu}
\otimes
\Ket{\tilde{z}_\nu'}\Bra{\tilde{z}_\nu}
\;.
\end{multline}
Inserting it into Eq.~\eqref{e:1-rdm} one arrives at the 1-RDM of the physical mode $z$
expressed
in the DVR approximation
\begin{multline} \label{e:rho1-tfd-mctdh}
{\rho}_{t|z}(z_\mu | z_{\mu'})= 
{\rm Tr}_{\tilde{z}}
\sum_{J}^{\notin (z,\tilde{z})} \sum_{J'}^{\notin (z,\tilde{z})}  A_J^*(t) A_{J'}(t) 
\delta_{J^{(z\tilde{z})},{J'}^{(z\tilde{z})}}
\sum_{\nu\nu' \in \tilde{z}} \times\\
C_{j_z, \mu_z'}^*(t)
C_{j_{\tilde{z}}, \nu_{\tilde{z}}'}^*(t)
C_{j_z',\mu_z}(t)
C_{j_{\tilde{z}}',\nu_{\tilde{z}}}(t) 
\Ket{z_\mu' \otimes \tilde{z}_\nu'}_{\rm iBT}\Bra{z_\mu \otimes \tilde{z}_\nu}
\;,
\end{multline}
where in this case
$\Ket{z_\mu' \otimes \tilde{z}_\nu'}_{\rm iBT}\Bra{z_\mu \otimes \tilde{z}_\nu} 
$
is the effective 2-dimensional DVR grid with correlations introduced solely by the
iBT. Note that, alternatively, another variant is possible,
\begin{multline}\label{e:rho1-tfd-mctdh-alt}
 \rho_{t|z}(z_\mu | z_{\mu'}) = 
\sum_{J}^{\notin (z,\tilde{z})} \sum_{J'}^{\notin (z,\tilde{z})}  A_J^*(t) A_{J'}(t) 
\delta_{J^{(z\tilde{z})},{J'}^{(z\tilde{z})}}
\times\\
 \sum_{\nu\in \tilde{z}}  
C_{j_z, a_\alpha^{(\beta)}(z_\mu, {\tilde{z}}_{\nu})}^*(t)
C_{j_{\tilde{z}}, b_\alpha^{(\beta)}(z_{\mu}, {\tilde{z}}_\nu)}^*(t)\times\\
C_{j_z',a_\alpha^{(\beta)}(z_{\mu'}, {\tilde{z}}_{\nu})}  (t)
C_{j_{\tilde{z}}',b_\alpha^{(\beta)}(z_{\mu'}, {\tilde{z}}_\nu)}  (t)
 \;,
\end{multline}
where this time we assumed the original DVR grids and time-dependent interpolations
of the 1-dimensional SPF coefficients. This is less costly to evaluate
and does not require modification of the original DVR grids used for the numerical simulations.

Nevertheless, expressions in Eq.~\eqref{e:rho1-tfd-mctdh-comb}, ~\eqref{e:rho1-tfd-mctdh} and ~\eqref{e:rho1-tfd-mctdh-alt}
are all very difficult to evaluate numerically
due to the 
the need of tensor contractions over 2-dimensional DVR quantities,
and a computationally feasible approximate approach is necessary
for larger systems and ML-MCTDH simulations.
In this work, we focus on the reduced 1-particle density that requires calculation
of the diagonal part of 2-RDM. In this case,
we implemented the calculation of the exact $\rho_z(z|z)$ from Eq.~\eqref{e:rho1-tfd-mctdh-alt}
by i) first extracting the diagonal part of the iBT 2-RDM (which is a correlation map between the real
and tilde mode), ii) two-dimensional interpolation of this map
by using the linear mappings from Eq.~\eqref{e:mappings}, and iii) tracing out the $\tilde{z}$ mode.
We found that such an approach can be implemented for  medium sized MCTDH wavefunctions,
as well as even ML-MCTDH wavefunctions that contain correlated wavefunction trees for a limited
amount of vibrational modes while the rest of the system adopts the Hartree approximation.
This minimizes the amount of the MCTDH configurations that can always be extracted from the ML-MCTDH simulations.
We anticipate that, except for such exceptional cases, in general the evaluation cost of Eq.~\eqref{e:rho1-tfd-mctdh-alt}
is prohibitive, both in terms of the CPU time and memory. Therefore, in the forthcoming sections
we discuss two approximate schemes.

\section{Approximations}
\subsection{Neglecting correlations between real and tilde modes}

In the special case where the real and tilde modes in the iBT formalism are uncorrelated,
the corresponding 2-RDM is a product of 1-RDMs of real and tilde mode, i.e.,
\begin{equation}
\label{e:app1}
{\gamma}^\beta_{t|z\tilde{z}}(z,\tilde{z} | z', \tilde{z}') = 
{\gamma}^\beta_{t|z}(z | z')\, {\gamma}^\beta_{t|\tilde{z}}(\tilde{z} | \tilde{z}') \;,
\end{equation}
leading to
\begin{multline}
\rho_{t|z}(z|z') = \int  
{\gamma}^\beta_{t|z}(a_{\alpha}^{(\beta)}(z , \tilde{z}) \big\vert  
                      a_{\alpha}^{(\beta)}(z',\tilde{z}))\times\\
{\gamma}^\beta_{t,z}(b_{\alpha}^{(\beta)}(z, \tilde{z}) \big\vert 
                      b_{\alpha}^{(\beta)}(z',\tilde{z})) \,{\rm d}\tilde{z} \;.
\end{multline}
The above equations are exact for harmonic systems and approximate
when correlations between the real and tilde modes are introduced (either directly or indirectly) by the anharmonicity.
It is expected though that at initial times the correlations
should be relatively small and Eq.~\eqref{e:app1} should be valid, at least approximately.
Note also that applying the special case 2-RDM from Eq.~\eqref{e:app1} to Eq.~\eqref{e:1-rdm}
will still correlate
the $z$ and $\tilde{z}$ modes in the physical system,
purely due to entanglement due to the thermal Bogoliubov transformation.

\subsection{Moment expansion}

Another strategy can utilize the fact that the moments of the physical mode position and momentum
are relatively easily calculable from the iBT formalism,
\begin{multline}
\label{e:moments}
M_{nm}(\beta;t)\equiv\left< z^np_z^m(t)\right> = 
\sum_{k=0}^n \sum_{l=0}^m
\binom{n}{k}
\binom{m}{l} \times\\
\cosh^{k+l}(\theta) 
\sinh^{n+m-k-l}(\theta) M_{n,m;n-k,m-l;{\rm iBT}}(\beta;t) \;,
\end{multline}
where
\begin{equation}
M_{i,j;k,l;{\rm iBT}}(\beta;t) \equiv \tBraKet{\Phi_t^\beta}{{z}^n  {p}_z^m {\tilde{z}}^{n-k} {\tilde{p}}_z^{m-l}}{\Phi_t^\beta} \;.
\end{equation}
\begin{equation}
\label{mn}
M_{n0}(\beta;t) \equiv \int {\rm d}z \, \rho_t(z) z^n \;.
\end{equation}

\subsubsection{Reduced 1-particle density}
First we seek a way to reconstruct $\rho^{(1)}(z)$ from a limited amount of its moments $M_{n0}$ for $n=0,1,\ldots,n_{\rm max}$
(the superscript (1) will be omitted from now on to simplify the notation).
We assume that probability density can be expressed in terms of Hermite polynomials
as follows:
\begin{equation}
\label{rho-approx-moments}
\rho(z) = \sum_{k=0}^{\infty} d_k H_k\left( \frac{z}{\sqrt{2} \sigma}\right) e^{-\frac{z^2}{2\sigma^2}} \;,
\end{equation}
with, for now, unknown expansion coefficients $d_k$. In the above equation,
the basis of the weighted orthogonal Hermite functions is centered around the origin and the zeroth-order function
is the normal distribution with variance of $\sigma^2$.

Now, using the fact that~\cite{NIST:DLMF}
\begin{equation}\label{xn}
z^n \equiv \frac{1}{2^n} \sum_{m=0}^{\lfloor n/2 \rfloor} \frac{(-n)_{2m}}{m!} H_{n-2m}(z) 
\end{equation}
one can substitute Eq.~\eqref{xn} into Eq.~\eqref{mn} and, noting the properties of Hermite polynomials,
the moments can be re-cast as
\begin{equation}
\label{mn-expl}
M_{n0} \equiv \sqrt{\pi} \sigma^{n+1} n! 2^{\frac{-n+1}{2}} 
\sum_{m=0}^{\lfloor n/2 \rfloor} \frac{2^{n-2m}}{m!} d_{n-2m} \;.
\end{equation}
In the above equations,
the Pochhammer symbol is defined as
\begin{equation}
 (s)_{n} \equiv \frac{\Gamma(s+n)}{\Gamma(s)}  \;.
\end{equation}
$\Gamma(s)$ is the Gamma function and $\lfloor n \rfloor$ denotes the floor function.
Thus, one can directly calculate the expansion coefficients from Eq.~\eqref{rho-approx-moments}
and the result is
\begin{equation}
\label{ak}
d_n \equiv \frac{M_{n0}}{\sqrt{\pi}\sigma^{n+1}n!2^{\frac{n+1}{2}}}
 -\sum_{m=1}^{\lfloor n/2 \rfloor} \frac{d_{n-2m}}{2^{2m}m!}  
 \;.
\end{equation}
They are simply given in terms of the raw moments and can be calculated recursively starting from the zeroth order.
Note also that one can also calculate the approximate momentum distribution $\rho(p_z)$
by evaluating the momentum moments $M_{0n}$ instead of the position moments $M_{n0}$.

\subsubsection{Algorithm for reduced 1-particle densities}
\label{sss:alg-1-dens}

Here we propose a simple algorithm to calculate the approximate reduced 1-particle
density from a finite series of its moments.
Since, as stated above, the orthogonal basis is centered at $z=0$ and also has a hyperparameter $\sigma$,
one should optimize the two hyperparameters of the orthogonal basis: $\sigma$ and the shift
along the $z$-axis, say $\mu$.
In effect, the first approximation of the distribution parameterically depends on them as follows:
\begin{multline}
\rho(z) \approx \rho^{(n_{\rm max})}(x;\sigma,\mu=0)  \\=
\sum_{k=0}^{n_{\rm max}} d_k(\sigma,\mu=0) H_k\left( \frac{z}{\sqrt{2} \sigma}\right) e^{-\frac{z^2}{2\sigma^2}} \;.
\end{multline}
We define a shift operator, $\mathcal{S}_\mu[f(x)]\equiv f(x-\mu)$.
Using the binomial theorem, the moments of the shifted distribution
are related to the unshifted distribution via
\begin{equation}
\label{mnmu}
M_{n0}(\mu) = \sum_{k=0}^n \frac{n!\mu^k}{k!(n-k)!}  M_{n-k,0}(\mu=0) \;.
\end{equation}
Then one can envisage the following algorithm:
\begin{enumerate}
 \item Compute all the moments $M_{n0}(\mu=0)$ for $n=0,1,\ldots,n_{\rm max}$ of the distribution to be found;
 \item Optimize $\sigma$ and $\mu$ hyperaparameters:
 \begin{enumerate}
    \item Initialize $\sigma = M_{2,0} - M_{1,0}^2$ and $\mu = -M_{1,0}$, where $M_{1,0}$ and $M_{2,0}$ are the moments
          of the unknown (unshifted) distribution;
    \item Iteratively converge $\sigma$ and $\mu$:
    \begin{itemize}
       \item Shift the distribution by $\mu$, i.e., calculate $M_{n0}(\mu)$ from Eq.~\eqref{mnmu};
       \item Calculate $d_k(\sigma,\mu)$ coefficients from Eq.~\eqref{ak} based on $M_{n0}(\mu)$ and a current $\sigma$. This defines the shifted
             distribution
             \begin{multline*}
              \qquad\qquad\mathcal{S}_\mu[\rho^{(n_{\rm max})}(z;\sigma,\mu)] = \\
               \sum_{k=0}^{n_{\rm max}} d_k(\sigma,\mu) H_k\left( \frac{z}{\sqrt{2} \sigma}\right) e^{-\frac{z^2}{2\sigma^2}} \;;
             \end{multline*}
       \item Check the convergence. If converged, exit the iterations.
             The optimal values are $\mu^*$ and $\sigma^*$.
    \end{itemize}
    \item Unshift the distribution by the optimal $\mu^*$ value,  i.e.,
             \begin{multline*}
              \qquad\mathcal{S}_{-\mu^*}\left[\mathcal{S}_{\mu^*}[\rho^{(n_{\rm max})}(z;\sigma^*,\mu^*)]\right] = \\
               \sum_{k=0}^{n_{\rm max}} a_k(\sigma^*,\mu^*) H_k\left( \frac{z+\mu^*}{\sqrt{2} \sigma}\right) e^{-\frac{(z+\mu^*)^2}{2\sigma^2}} \;.
             \end{multline*}
    The above result approximates the original distribution, i.e.,
             \begin{equation*}
             \,\qquad\rho(z)\approx \mathcal{S}_{-\mu^*}\left[\mathcal{S}_{\mu^*}[\rho^{(n_{\rm max})}(z;\sigma^*,\mu^*)]\right] \;.
             \end{equation*}
 \end{enumerate}
\end{enumerate}
As the criterion for the optimization, we minimize the absolute negative norm
of the approximate distribution according to
\begin{equation}
\rho^{(n_{\rm max})}(z;\sigma^*,\mu^*) = \min_{\mu,\sigma} \int {\rm d}z\, \mathcal{F}[\rho^{(1)}(z) ] \;,
\end{equation}
where $\mathcal{F}[f(x)] = -f(x)$ if $f(x) < 0$ and 0 otherwise.

\begin{figure*}
\centering
\includegraphics[width=0.8\linewidth]{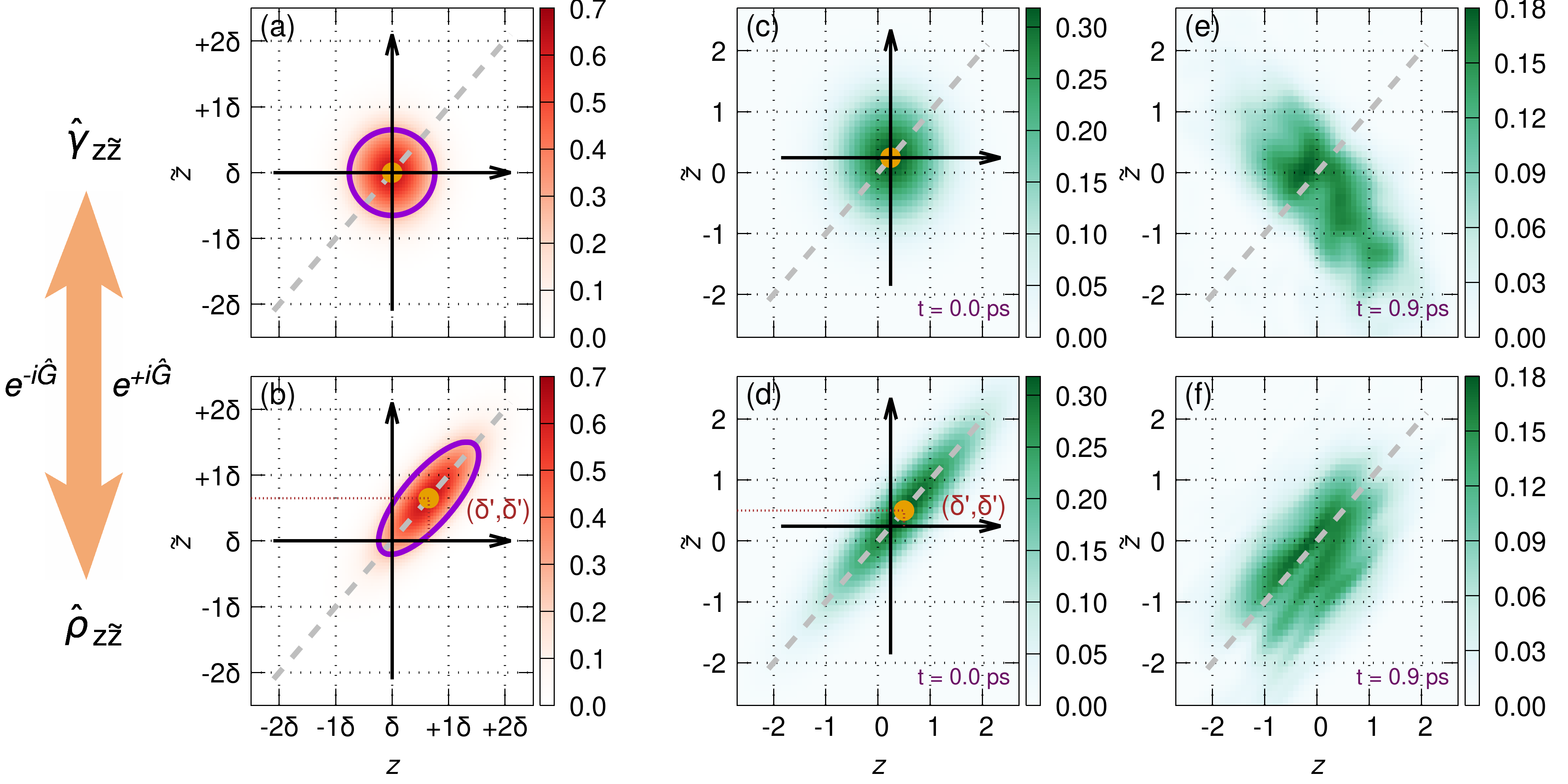}
\caption{Two-mode squeezing effect on the diagonal part of the 2-RDM under the iBT formalism.
(a) Gaussian 2-RDM; 
(b) Analytical result of the iBT-transformed 2-RDM from (a);
(c-f) actual diagonal 2-RDMs from MCTDH calculations for the weakly quartic
oscillator of Eq.\ (\ref{e:example-h0_1}) at different times. 
$\delta$ denotes the location of the shifted initial condition in the iBT position space
and $\delta'$ is the transformed initial condition in the physical position space.
Black axes in panels (a-d) help to emphasize the shift.
In (c-d), the state at $t = 0$ is shown, while (e-f) shows the state at 0.9 ps
in the complementary representations. 
}
\label{fig-0}
\end{figure*}

\subsubsection{Wigner distribution}

Now we consider the Wigner distribution $W(z,p_z)$ whose moments are given as
\begin{equation}
\label{wmn}
M_{nm} \equiv \iint {\rm d}z \,{\rm d}p_z \,  W(z,p_z) z^n p_z^m \;.
\end{equation}
Following the strategy from the previous section we expand it as
\begin{multline}
\label{wig}
W(z,p_z) = \sum_{k=0}^{\infty} \sum_{l=0}^{\infty} d_{kl} \times\\
         H_k\left( \frac{z  }{\sqrt{2} \sigma_z}\right) 
         H_l\left( \frac{p_z}{\sqrt{2} \sigma_p}\right) 
         e^{-\frac{z^2  }{2\sigma_x^2}} 
         e^{-\frac{p_z^2}{2\sigma_p^2}} 
         \;,
\end{multline}
where now $\sigma_x$ and $\sigma_p$ are hyperparameters associated with the width of the position and momentum distribution, respectively.
Note that the above equation contains correlations between $x$ and $p$ solely in the expansion coefficients
$d_{kl}$. The expression from Eq.~\eqref{wig} is different from the previously used moment expansion of the Wigner
distribution\cite{Hughes.etal.JCP.2009} in that the expansion is done for both position and momentum separately,
and not of the momentum only. This results in slightly simpler to evaluate analytical expressions,
although the approaches are formally equivalent.
Applying Eq.~\eqref{xn} to $z^n$ and $p_z^m$, the Wigner moments can be re-expressed as
\begin{multline}
\label{wmn-expl}
M_{nm} \equiv \pi \sigma_x^{n+1}\sigma_p^{m+1} n!m! 2^{\frac{2-n-m}{2}} \times\\
\sum_{k=0}^{\lfloor n/2 \rfloor} \sum_{l=0}^{\lfloor m/2 \rfloor}  \frac{2^{n+m-2k-2l}}{k!l!} d_{n-2k,m-2l} \;.
\end{multline}
Thus, the expansion coefficients are given by the following recurrence relationship
\begin{equation}
\label{akl}
d_{nm} \equiv \frac{M_{nm}}{s_{nm}2^{n+m}}
 - 
  \sum_{k}^{\lfloor n/2 \rfloor} \sum_{l}^{\lfloor m/2 \rfloor} \frac{d_{n-2k,m-2l} }{2^{2k+2l}k!l!} \Bigg|_{kl\neq 00}
 \;,
\end{equation}
where $s_{nm}$ is the multiplicative prefactor before the summation terms in Eq.~\eqref{wmn-expl}.

\subsubsection{Algorithm for Wigner and higher-dimensional distributions}

The one-dimensional algorithm from Sec.~\ref{sss:alg-1-dens} can be easily extended to two-dimensional
Wigner distributions. In this case,
the optimization space consists of four hyperparameters
$\{\mu_z, \mu_p, \sigma_z, \sigma_p\}$, which could be done by minimizing the differences between marginals
$\left| \rho(z) - \int {\rm d}p_z W(z,p_z) \right|$ and
$\left| \rho(p) - \int {\rm d}z   W(z,p_z) \right|$
where $\rho(z)$ and $\rho(p)$ are the approximants of the position and momentum distribution
obtained from the 1-dimensional algorithm.

We note that, in principle, approximants to other two-dimensional distributions
(e.g., two-particle distribution between two different physical modes),
as well as general higher-dimensional
distributions can be extracted in similar way by the Hermite function expansions
of the ${x}^n$ or ${p}_x^m$ terms. The method described here
can also be used in general quantum dynamics simulations.
This would however require relatively large
number of moments that grows exponentially with dimension of the distribution.
In this work, we focus on one-dimensional distributions in position space as a proof of concept
applied in the iBT TFD context.

\section{Demonstrative Example: Thermalized Anharmonic Oscillator}
\label{sec:results}

To demonstrate the use of the exact (Eq.~\eqref{e:rho1-tfd-mctdh}) 
and approximate (Eqs.~\eqref{rho-approx-moments}, as well as
Eq.~\eqref{e:app1} and \eqref{e:1-dens}) expressions
in calculations of the reduced 1-particle densities,
we consider a model of a quartic anharmonic oscillator 
\begin{equation} 
\label{e:example-h0_1}
 {H} = \frac{\omega_z}{2} \left({z}^2 + {p}_z^2\right) 
 + a_3{z}^3 + a_4{z}^4
 \;,
\end{equation}
where $\omega_z$ is the harmonic frequency of 200~cm$^{-1}$
whereas $a_3$ and $a_4$ are the cubic and quartic anharmonic constants, 
equal to $7.35\cdot 10^{-5}$ and $7.35\cdot 10^{-6}$~a.u., respectively.
The iBT Hamiltonian thermalized using the HO approximation
centered at the origin
can be readily found to be
\begin{multline} 
\label{e:example-h0_2}
 {K}(\beta) = \frac{\omega_z}{2} \left({z}^2 + {p}_z^2 - {\tilde{z}}^2 - {\tilde{p}}_z^2\right) \\
 + a_3\left(\cosh\theta {z} + \sinh\theta {\tilde{z}}\right)^3 
 + a_4\left(\cosh\theta {z} + \sinh\theta {\tilde{z}}\right)^4
 \;.
\end{multline}
We now consider a displaced thermal state and initialize MCTDH simulations with an initial condition
$
\BraKet{z,\tilde{z}}{\Phi^\beta(t=0)} \propto
 e^{-\frac{1}{2}(       z -z(0))^2} 
 e^{-\frac{1}{2}(\tilde{z}-z(0))^2} 
$ with $z(0)=0.5e^{-\theta(\beta)}$~a.u. which ensures that the initial physical mode expectation value
is temperature independent and equal to 0.5~a.u.
12 SPFs are used for the initially separable particles $z$ and $\tilde{z}$,
which are represented on a Fast Fourier Transform (FFT) DVR grid. The simulations are converged
within a natural orbital population threshold of 0.5{\%} at 1~ps.

\begin{figure}
\centering
\includegraphics[width=1.0\linewidth]{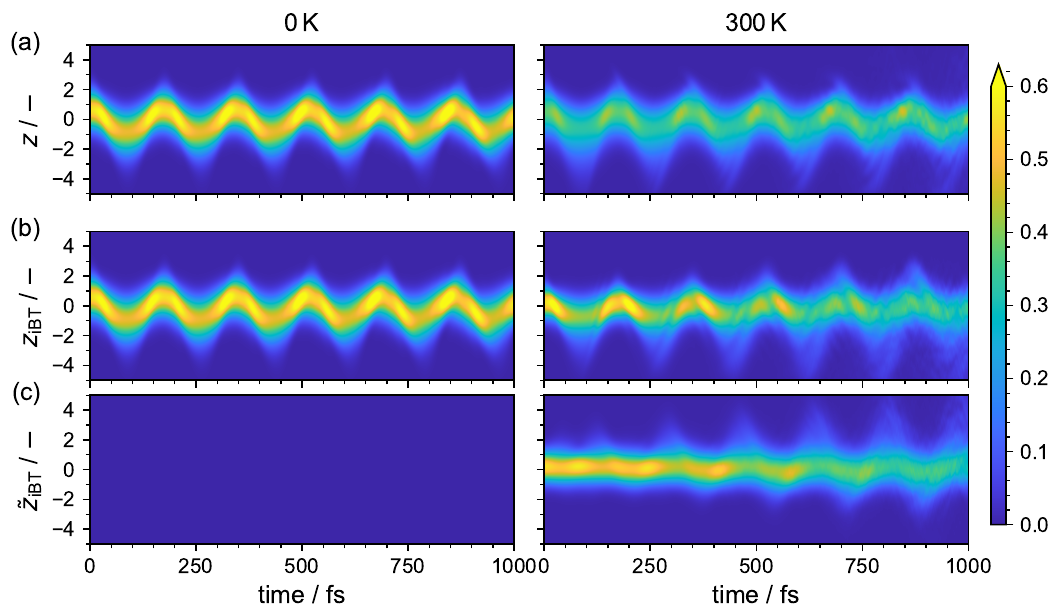}
\caption{Time evolution of the reduced 1-particle density of the anharmonic oscillator
at $T =$ 0 K (l.h.s.) and $T =$ 300K (r.h.s.). The 1-particle densities $\rho^{(1)}_{\rm iBT}(z)$ and $\rho^{(1)}_{\rm iBT}(\tilde{z})$ are also shown
for comparison. At non-zero temperature, the latter do not have a physical meaning.
}
\label{fig-1}
\end{figure}
In Fig.~\ref{fig-0} we show the effect of the iBT on the diagonal part of the 2-RDM
that results in two-mode squeezing\cite{Caves.Schumaker.PRA.1985,Barnett85} of the real and tilde modes.
In the limit of an uncorrelated iBT 2-RDM (e.g., at the initial condition),
one can model it via a Gaussian function static initial condition
$\gamma^\beta_{z\tilde{z}}(z|\tilde{z}) = (2\pi\sigma^2)^{-1} {\rm exp}\left(
-\{(z-\delta)^2 + (\tilde{z}-\delta)^2\}/{(2\sigma^2)}
\right)$ shown in Fig.~\ref{fig-0}(a). Its uncertainty ellipse 
is circular due to the lack of correlation between $z$ and $\tilde{z}$ positions.
The iBT with the HO shifted in position space by $\Delta$ yields
$\rho_{z\tilde{z}}(z|\tilde{z}) = (2\pi\sigma^2)^{-1} {\rm exp}\left(
-\{(a_\Delta^{(\beta)}(z,\tilde{z})-\delta)^2 + (b_\Delta^{(\beta)}(z,\tilde{z})-\delta)^2\}/{(2\sigma^2)}
\right)$
with its initial position shifted according to the mappings from Eq.~\eqref{e:mappings}
to $\delta' = e^{\theta}\left[\delta-\Delta(1-e^{-\theta})\right]$,
shown in Fig.~\ref{fig-0}(b).
Due to the symplectic nature of the transformation, 
the phase space area in the position space of both modes is preserved,
hence the characteristic elongation 
of the uncertainty ellipse along the 45$^{\circ}$ axis.~\cite{wang2024structured,Glorieux_2023,Bello.2021}
Note that in a special case when the shift of the HO potential is equivalent to the initial condition,
$\Delta = \delta$, 
the initial condition location is invariant under the iBT, i.e., $\delta'=\delta=\Delta$,
regardless of the squeezing strength measured by $\theta$.

The above discussed analytical model can be subsequently compared to the actual diagonal 2-RDMs 
obtained from MCTDH simulations of the anharmonic oscillator system i) at the initial condition
(Fig.~\ref{fig-0}(c-d)), and ii) at 0.9~ps where correlations between the modes
appear in the iBT space due to the anharmonicity of the potential (Fig.~\ref{fig-0}(e-f)).
The squeezing at the initial condition 
distorts the iBT 2-RDMs much in the same way as seen in the analytical model.
Since in this particular case $\Delta = 0$, the non-zero shift occurs as
$\delta' = e^{\theta}\delta$.
It is also interesting to see how the more complex 2-RDM at later time gets stretched along
the 45$^{\circ}$ axis, with each of the peaks in the iBT space 
being stretched along their local axes with a simultaneous anti-squeezing
in the orthogonal direction, bringing the features closer to the central squeezing axis. 
This emphasizes the non-trivial effects 
on the physical mode properties including the one-particle densities
due to the two-mode squeezing effect induced by the iBT in a general setting with 
a combination of anharmonic and thermal correlations.

Fig.~\ref{fig-1} shows the time evolution of the reduced 1-particle density
at 0K and 300K. The effect of temperature is clearly visible in the broadening
of the distribution of the physical mode. Also, the reduced 1-particle densities
of the real and tilde modes associated with the iBT wavefunctions
are quite different from the actual physical mode distribution.

\begin{figure}
\centering
\includegraphics[width=1.0\linewidth]{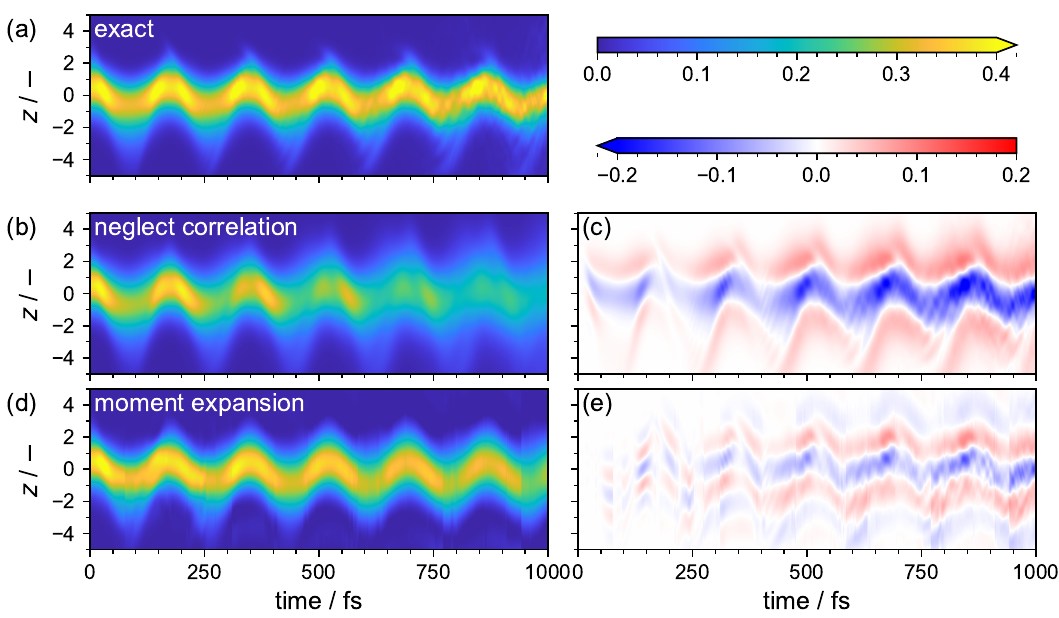}
\caption{Exact and approximate reduced 1-particle density of the anharmonic
oscillator. (a) Exact time evolution, (b) uncorrelated approximation, (c)
deviation of the uncorrelated approximation from the exact result,
(d) moment approximation, (e) deviation of the moment approximation from the
exact result.}
\label{fig-2}
\end{figure}

\begin{figure*}[hbt!]
\centering
\includegraphics[width=0.7\linewidth]{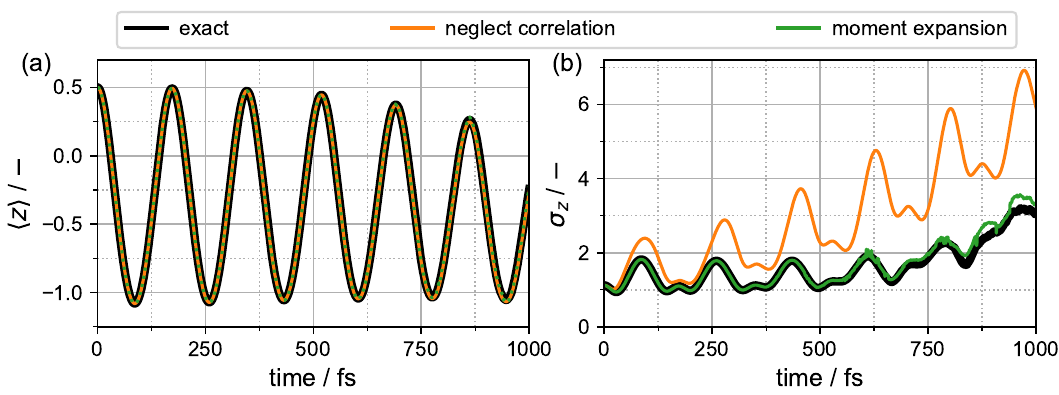}
\caption{First moment (left) and variance (right) of the physical mode
calculated with the two approximations discussed in the main text, i.e., the
uncorrelated approximation and the moment approximation.
}
\label{fig-3}
\end{figure*}

We applied the two approximations introduced above, i) the uncorrelated
approximation, and ii) the moment expansion up to 15th order, and the results are shown 
in Fig.~\ref{fig-2} as compared to the reference (exact) calculation
from the ML-MCTDH 2-RDM.
As expected, the uncorrelated approximation is close to exact at earlier times
and the errors accumulate at later times.
The moment expansion method works much better qualitatively reconstructing the physical
mode density. We set the threshold of 0.0001 of negative norm and we found that
starting from initial 20th order, the algorithm typically finished at orders between 6-15th 
depending on the time step.

Interestingly, the first moments of the mode densities are exactly reproduced
by both approximate schemes even when thermal correlations are completely neglected, 
as shown in the left panel of Fig.~\ref{fig-3}.
Only at the level of (at least) the variance (second cumulant) the differences are visible
(right panel of Fig.~\ref{fig-3}), with the moment expansion faithfully reconstructing
the exact wavepacket width. Neglecting thermal correlations leads to the loss of information over time
which manifests in an overly broadened wavepacket. Only at very early times below 50~fs
the width agrees with the exact calculation, which indicates
that the thermal correlations build up quite quickly even for the weakly anharmonic oscillator.

\section{Conclusions}
\label{sec:conclusions}

In this work we discussed the calculation of the 1-RDM, reduced 1-particle density and Wigner distributions
under the iBT formalism of TFD. We found that it is possible to calculate the 1-RDM from the exact expression
for relatively low-dimensional vibrational wavefunctions using the ML-MCTDH framework. The approximate moment expansion method
can be used to obtain a qualitative picture of the thermalized particles. In
principle, this approach can be easily generalized to higher-dimensional phase-space distributions.
We anticipate that in the case of ML-MCTDH simulations, the recurrence structure of the wavefunction
can be further employed in conjunction with Eq.~\eqref{e:1-rdm} to obtain exact distributions
that are less expensive to evaluate as compared with the MCTDH form of Eq.~\eqref{e:rho1-tfd-mctdh}
shown in this work. 

Recently, rigorous expressions for the reduced density matrix elements in the
quantum thermofield-theoretic setting were found using Feynman path integrals
by K{\"a}ding and Pitschmann.\cite{Kading.Pitschmann.PhysRevD.2023} Their
work, carried out in a high-energy and quantum field theory (QFT) context,
provides explicit formulas for time-dependent reduced density operators in
open systems directly within the doubled Hilbert space formalism, and
naturally incorporates the iBT through the definition of thermal creation and
annihilation operators. While these results establish an elegant and general
analytical foundation for reduced density matrices in TFD from the QFT
perspective, it is not straightforward to transfer these results to low-energy problems
that are typically encountered in chemistry. Here, the focus is on
multi-configurational vibrational and electronic wavefunctions and tractable
reduced particle distributions.
In the present work, we have therefore concentrated 
on this regime, with the aim of developing a practicable formulation of reduced 
1-particle densities, Wigner and higher-order distributions within the 
iBT-TFD representation.

\begin{acknowledgments}
Financial support from the Leverhulme Trust Research Project Grant
PRG-2023-078 is gratefully acknowledged.
\end{acknowledgments}

\section*{Data Availability Statement}
The data that support the findings of this study are available from the corresponding author upon reasonable request.

\appendix

\section{Bogoliubov transformation for bosonic and fermionic systems}
\label{a:bt}

Since the transformations employed in TFD are a special case of Bogoliubov
transformations (BTs) we briefly outline the general properties of
BTs for bosonic and fermionic systems in the following, focusing on
their two-mode version which is of direct relevance for our purposes.
The key feature of BTs is the mixing of creation and annihilation
operators pertaining to pairs of Hilbert spaces while conforming to
the relevant (anti-)commutation rules. We then turn to the specific
properties of the double Hilbert space of TFD.

For a pair of annihilation operators ${a}_{1}$ and ${a}_{2}$ (and their
adjoint) let us introduce the doublets 
\[
\boldsymbol{\Psi}=\begin{pmatrix}{a}_{1}\\
{a}_{2}^{\dagger}
\end{pmatrix}\;,\ \ \bar{\boldsymbol{\Psi}}=\begin{pmatrix}{a}_{1}^{\dagger}\\
{a}_{2}
\end{pmatrix} \,.
\]
The two-mode Bogoliubov transformation is defined as 
\[
\boldsymbol{\Psi}'=U_{\sigma}\boldsymbol{\Psi}\;,\ \ \bar{\boldsymbol{\Psi}}'=U_{\sigma}^{*}\bar{\boldsymbol{\Psi}}\;,
\]
where $U_{\sigma}$ is a $2\times2$ non-singular matrix, required
to preserve the (anti)commutation rules, i.e. $[{a}_{i},{a}_{j}^{\dagger}]_{\sigma}=\delta_{ij}$
and $[{a}_{i},{a}_{j}]_{\sigma}=[{a}_{i}^{\dagger},{a}_{j}^{\dagger}]_{\sigma}=0$
if $\sigma=\pm1$ is used for fermions and bosons, respectively. The
latter rules are conveniently encoded in a matrix form as follows
\begin{multline}
{M}_{\sigma}:=\boldsymbol{\Psi}\boldsymbol{\Psi}^{\dagger}+\sigma\left(\bar{\boldsymbol{\Psi}}\bar{\boldsymbol{\Psi}}^{\dagger}\right)^{t}
\\
=\left[\begin{array}{cc}
[{a}_{1},{a}_{1}^{\dagger}]_{\sigma} & [{a}_{1},{a}_{2}]_{\sigma}\\{}
[{a}_{2}^{\dagger},{a}_{1}^{\dagger}]_{\sigma} & [{a}_{2}^{\dagger},{a}_{2}]_{\sigma}
\end{array}\right]=
\left[\begin{array}{cc}
1 & 0\\
0 & \sigma
\end{array}\right]\;.
\end{multline}
Here, the adjoint ($\dagger$) is defined to transpose ($t$) the
arrays and take the Hermitean conjugate of their elements.
Since both terms in this expression transform congruently under a
Bogoliubov map $\boldsymbol{\Psi}\rightarrow {U}_{\sigma}\boldsymbol{\Psi}$,
we require
\begin{equation}
{U}_{\sigma}{M}_{\sigma}{U}_{\sigma}^{\dagger}={M}_{\sigma} \;.\label{eq:isometry}
\end{equation}
This identifies $U_{+1}\in SU(2)$ and $U_{-1}\in SU(1,1)$, after
removing an irrelevant global phase. The resulting BT matrices take
the general form
\begin{equation}
{U}_{\sigma}(u,v)=\left[\begin{array}{cc}
u & v\\
-\sigma v^{*} & u^{*}
\end{array}\right]\label{eq:U form}
\end{equation}
with $|u|^{2}+\sigma|v|^{2}=1$, and satisfy ${U}_{\sigma}{M}_{\sigma}{U}_{\sigma}^{\dagger}={U}_{\sigma}^{\dagger}{M}_{\sigma}{U}_{\sigma}$.
This immediately shows that the quadratic form
\begin{equation}
\boldsymbol{\Psi}^{\dagger}{M}_{\sigma}\boldsymbol{\Psi}={a}_{1}^{\dagger}{a}_{1}+\sigma {a}_{2}{a}_{2}^{\dagger}\label{eq:invariant}
\end{equation}
is invariant under a Bogoliubov transformation. Furthermore, gauging
away the remaining phase factors, ${U}_{\sigma}$ is conveniently parametrized
as 
\begin{align*}
{U}_{+1}(\theta)&=\left[\begin{array}{cc}
\cos(\theta) & -\sin(\theta)\\
\sin(\theta) & \cos(\theta)
\end{array}\right]\;,\ \ \ \\
{U}_{-1}(\theta)&=\left[\begin{array}{cc}
\cosh(\theta) & -\sinh(\theta)\\
-\sinh(\theta) & \cosh(\theta)
\end{array}\right]\;.
\end{align*}
The angle $\theta$ is a true rotation angle for fermions\footnote{In the fermionic ($SU(2)$) case this amounts to use the Euler parametrization
$(\varphi,\theta,\psi)$ and keep only the rotation around $y$ by
the angle $\theta$.} and a hyperbolic \textquotedblleft squeeze\textquotedblright{} parameter
for bosons, and can be fixed to make the BT useful for the problem
at hand. The BT is then implemented by a unitary operator
\[
{a}'_{i}=e^{-{\rm i}\theta {F}}{a}_{i}e^{+{\rm i}\theta {F}}\;,
\]
which is generated by 
\begin{equation}
{F}={F}^{\dagger}={\rm i}\left({a}_{1}^{\dagger}{a}_{2}^{\dagger}-{a}_{2}{a}_{1}\right)\label{eq:generator}
\end{equation}
for both fermions and bosons. Indeed, one checks that 
\[
\frac{d{a}'_{1}}{d\theta}=-{\rm i}[{F},{a}'_{1}]_{-}=-{a}'{}_{2}^{\dagger}\ \ \frac{d{a}'{}_{2}^{\dagger}}{d\theta}=-{\rm i}[{F},{a}'{}_{2}^{\dagger}]_{-}=+{a}'_{1}
\]
integrate to the Bogoliubov form above. 

For the problem described in the main text, where ${a}_{1}={a}$ and ${a}_{2}={\tilde{a}}$
are turned into ${a}({\theta})$ and ${\tilde{a}}({\theta})$, respectively,
the Bogoliubov - transformed operators are required to annihilate
the thermal state in the thermofield representation, i.e. the angle
$\theta$ is fixed by the condition 
\begin{equation}
{a}({\theta})\ket{{0}(\beta)}={\tilde{a}}({\theta})\ket{{0}(\beta)}=0 \;,\label{eq:annihilation for the thermal vacuum}
\end{equation}
leading to the equations
\[
\exp\left(-\frac{\beta\hbar\omega}{2}\right)=\left\{ \begin{array}{ccc}
\tan(\theta) &  & \text{for }\sigma=+1\\
\\\tanh(\theta) &  & \text{for }\sigma=-1
\end{array}\right.
\]
which define $\theta\equiv\theta(\beta)=\Theta_{\beta}$ for $\sigma=\pm1$.
In first quantization, it is often more convenient to treat the Bogoliubov
map as a change of picture. Specifically, since by construction $\Theta_{\beta}$
is such that 
\[
{a}\left(e^{+{\rm i}\Theta_{\beta}{F}}\ket{{0}(\beta)}\right)={\tilde{a}}\left(e^{+{\rm i}\Theta_{\beta}{F}}\ket{{0}(\beta)}\right)=0
\]
one conveniently performs the following unitary transformation of
the thermofield dynamics from the double space description (DS) to
the Bogoliubov one ($\Theta_{\beta}$)
\[
\Ket{\Psi_{t}}_{\Theta_{\beta}}:=e^{+{\rm i}\Theta_{\beta}{F}}\Ket{\Psi_{t}}_{\text{DS}}\;,\ \ \ A_{\Theta_{\beta}}:=e^{{\rm i}\Theta_{\beta}{F}}A_{\text{DS}}e^{-{\rm i}\Theta_{\beta}{F}}\;.
\]
Here, the new operators ${A}({\Theta_{\beta}})$'s are seen to transform
according to the \emph{inverse} Bogoliubov transformation (iBT), in
such a way that the annihilation operators for the thermal vacuum
$\ket{{0}(\beta)}$ defined in Eq. \ref{eq:annihilation for the thermal vacuum}
are mapped back to the annihilation operators of the bare vacuum $\Ket{\mathbf{0}}=\Ket{0,\tilde{0}}$
\[
e^{{\rm i}\Theta_{\beta}{F}}{a}({\Theta_{\beta}})e^{-{\rm i}\Theta_{\beta}{F}}\equiv {a},\ \ \ e^{{\rm i}\Theta_{\beta}{F}}{\tilde{a}}({\Theta_{\beta}})e^{-{\rm i}\Theta_{\beta}{F}}\equiv{\tilde{a}}\;.
\]
and the equilibrium state is represented by $\Ket{\mathbf{0}}$ at
any temperature.

\section{Bogoliubov transformation with displaced bosonic creation and annihilation operators}
\label{a:displaced-bogoliubov}

Here we consider bosonic systems. We rewrite the shifted bosonic Bogoliubov transformation in the following way:
\begin{equation}
  e^{-\theta ({\tilde{a}}_\alpha^\dagger {a}_\alpha^\dagger - {\tilde{a}}_\alpha{a}_\alpha)} 
 \equiv 
 e^{{A}-{B}} \;,
\end{equation}
where
\begin{subequations}
\begin{align}
 {A} &= -\theta \left\{ {\tilde{a}}^\dagger {a}^\dagger - {\tilde{a}} {a} \right\} \;,\\
 {B} &= -\theta \left\{ \alpha ({a}^\dagger - {\tilde{a}} ) + \alpha^* ({\tilde{a}}^\dagger - {a}) \right\} \;.
\end{align}
\end{subequations}
The commutator yields
$[{A}, {B}] = -\theta{B} $
which implies that $e^{{A}-{B}} = e^{{A}} e^{g(\theta){B}}$.
The solution for the function $g$ is~\cite{van-brunt.visser.JPA.2015}
\begin{equation}
 g(\theta) = \frac{1-e^{\theta}}{\theta} \;,
\end{equation}
which gives the relationship in Eq.~\eqref{e:bogo-rela-b}.
Now, by the Maclaurin expansion of the displacement operator
one can show that the transformation of the displacement operator
under the unshifted Bogoliubov transformation is:
\begin{subequations}\label{disp-bogo}
\begin{align}
 e^{{\rm i}{G}} {D}(\alpha) e^{-{\rm i}{G}} &=
 {       D }( \alpha \cosh{\theta})
 {\tilde{D}}(-\alpha \sinh{\theta}) \;, \\
 e^{{\rm i}{G}} {\tilde{D}}(\alpha) e^{-{\rm i}{G}} &=
 {\tilde{D}}( \alpha \cosh{\theta})
 {      {D}}(-\alpha \sinh{\theta}) \;.
\end{align}
\end{subequations}
Using this result in conjunction with Eq.~\eqref{e:bogo-rela-b} one arrives to
Eq.~\eqref{e:bogo-rela-a}. 

It is worth commenting on the practical implications
of Eq.~\eqref{e:bogo-rela-a}. Applying it to the iBT Hamiltonian from
Eq.\ \eqref{eq:BT H}
is equivalent to i) first, applying the unshifted HO Bogoliubov transformation,
ii) subsequently applying the temperature-dependent displacements
of position and momentum operators according to
\begin{subequations}
\label{e:disp-prescription}
\begin{align}
 {       z }       &\longrightarrow  {       z }   - \Delta_z    (1 - e^{-\theta(\beta)}) \;, \\
 {\tilde{z}}       &\longrightarrow  {\tilde{z}}   - \Delta_z    (1 - e^{-\theta(\beta)}) \;, \\
 {       p }_z     &\longrightarrow  {       p }_z - \Delta_{p_z}(1 - e^{-\theta(\beta)}) \;, \\
 {\tilde{p}}_z     &\longrightarrow  {\tilde{p}}_z - \Delta_{p_z}(1 - e^{-\theta(\beta)}) \;,
\end{align}
\end{subequations}
for $\Delta_{p_z}\equiv\sqrt{2}\Im{\alpha}$.
Note that alternatively one can always shift
the coordinate system origin such that it coincides with
the thermal ground state minimum and shifted and unshifted
Bogoliubov transformations become equal.
We prefer using the generalized version that includes displacements
in coordinate space
because it simplifies implementing the dynamical setup in the model systems studied in this work.
We also note that the expressions given here translate to the coordinate-space expressions 
of Ref.~\citenum{BBMB-1-2025} if only a coordinate shift is considered.

\bibliography{refs}

\begin{thebibliography}{75}%
\makeatletter
\providecommand \@ifxundefined [1]{%
 \@ifx{#1\undefined}
}%
\providecommand \@ifnum [1]{%
 \ifnum #1\expandafter \@firstoftwo
 \else \expandafter \@secondoftwo
 \fi
}%
\providecommand \@ifx [1]{%
 \ifx #1\expandafter \@firstoftwo
 \else \expandafter \@secondoftwo
 \fi
}%
\providecommand \natexlab [1]{#1}%
\providecommand \enquote  [1]{``#1''}%
\providecommand \bibnamefont  [1]{#1}%
\providecommand \bibfnamefont [1]{#1}%
\providecommand \citenamefont [1]{#1}%
\providecommand \href@noop [0]{\@secondoftwo}%
\providecommand \href [0]{\begingroup \@sanitize@url \@href}%
\providecommand \@href[1]{\@@startlink{#1}\@@href}%
\providecommand \@@href[1]{\endgroup#1\@@endlink}%
\providecommand \@sanitize@url [0]{\catcode `\\12\catcode `\$12\catcode
  `\&12\catcode `\#12\catcode `\^12\catcode `\_12\catcode `\%12\relax}%
\providecommand \@@startlink[1]{}%
\providecommand \@@endlink[0]{}%
\providecommand \url  [0]{\begingroup\@sanitize@url \@url }%
\providecommand \@url [1]{\endgroup\@href {#1}{\urlprefix }}%
\providecommand \urlprefix  [0]{URL }%
\providecommand \Eprint [0]{\href }%
\providecommand \doibase [0]{https://doi.org/}%
\providecommand \selectlanguage [0]{\@gobble}%
\providecommand \bibinfo  [0]{\@secondoftwo}%
\providecommand \bibfield  [0]{\@secondoftwo}%
\providecommand \translation [1]{[#1]}%
\providecommand \BibitemOpen [0]{}%
\providecommand \bibitemStop [0]{}%
\providecommand \bibitemNoStop [0]{.\EOS\space}%
\providecommand \EOS [0]{\spacefactor3000\relax}%
\providecommand \BibitemShut  [1]{\csname bibitem#1\endcsname}%
\let\auto@bib@innerbib\@empty
\bibitem [{\citenamefont {Takahashi}\ and\ \citenamefont
  {Umezawa}(1975)}]{Takahashi.Umezawa.Coll.1975}%
  \BibitemOpen
  \bibfield  {author} {\bibinfo {author} {\bibfnamefont {Y.}~\bibnamefont
  {Takahashi}}\ and\ \bibinfo {author} {\bibfnamefont {H.}~\bibnamefont
  {Umezawa}},\ }\href@noop {} {\bibfield  {journal} {\bibinfo  {journal}
  {Collect. Phenom.}\ }\textbf {\bibinfo {volume} {2}},\ \bibinfo {pages} {55}
  (\bibinfo {year} {1975})}\BibitemShut {NoStop}%
\bibitem [{\citenamefont {Umezawa}, \citenamefont {Matsumoto},\ and\
  \citenamefont {Tachiki}(1982)}]{umezawa1982thermo}%
  \BibitemOpen
  \bibfield  {author} {\bibinfo {author} {\bibfnamefont {H.}~\bibnamefont
  {Umezawa}}, \bibinfo {author} {\bibfnamefont {H.}~\bibnamefont {Matsumoto}},\
  and\ \bibinfo {author} {\bibfnamefont {M.}~\bibnamefont {Tachiki}},\
  }\href@noop {} {\emph {\bibinfo {title} {Thermo field dynamics and condensed
  states}}}\ (\bibinfo {year} {1982})\BibitemShut {NoStop}%
\bibitem [{\citenamefont {Ojima}(1981)}]{Ojima.AnP.1981}%
  \BibitemOpen
  \bibfield  {author} {\bibinfo {author} {\bibfnamefont {I.}~\bibnamefont
  {Ojima}},\ }\bibfield  {title} {\enquote {\bibinfo {title} {{Gauge fields at
  finite temperatures — “Thermo field dynamics” and the KMS condition and
  their extension to gauge theories}},}\ }\href@noop {} {\bibfield  {journal}
  {\bibinfo  {journal} {Ann. Phys.}\ }\textbf {\bibinfo {volume} {137}},\
  \bibinfo {pages} {1--32} (\bibinfo {year} {1981})}\BibitemShut {NoStop}%
\bibitem [{\citenamefont {Takahashi}\ and\ \citenamefont
  {Umezawa}(1996)}]{Takahashi.Umezawa.IJMPB.1996}%
  \BibitemOpen
  \bibfield  {author} {\bibinfo {author} {\bibfnamefont {Y.}~\bibnamefont
  {Takahashi}}\ and\ \bibinfo {author} {\bibfnamefont {H.}~\bibnamefont
  {Umezawa}},\ }\bibfield  {title} {\enquote {\bibinfo {title} {Thermo field
  dynamics},}\ }\href@noop {} {\bibfield  {journal} {\bibinfo  {journal} {Int.
  J. Mod. Phys. B}\ }\textbf {\bibinfo {volume} {10}},\ \bibinfo {pages}
  {1755--1805} (\bibinfo {year} {1996})}\BibitemShut {NoStop}%
\bibitem [{\citenamefont {Arimitsu}\ and\ \citenamefont
  {Umezawa}(1987)}]{Arimitsu.Umezawa.PTP.1987}%
  \BibitemOpen
  \bibfield  {author} {\bibinfo {author} {\bibfnamefont {T.}~\bibnamefont
  {Arimitsu}}\ and\ \bibinfo {author} {\bibfnamefont {H.}~\bibnamefont
  {Umezawa}},\ }\bibfield  {title} {\enquote {\bibinfo {title}
  {{{Non-Equilibrium Thermo Field Dynamics}}},}\ }\href@noop {} {\bibfield
  {journal} {\bibinfo  {journal} {Prog. Theor. Phys.}\ }\textbf {\bibinfo
  {volume} {77}},\ \bibinfo {pages} {32--52} (\bibinfo {year}
  {1987})}\BibitemShut {NoStop}%
\bibitem [{\citenamefont {Arimitsu}, \citenamefont {Guida},\ and\ \citenamefont
  {Umezawa}(1988)}]{Arimitsu.Guida.Umezawa.PhysA.1988}%
  \BibitemOpen
  \bibfield  {author} {\bibinfo {author} {\bibfnamefont {T.}~\bibnamefont
  {Arimitsu}}, \bibinfo {author} {\bibfnamefont {M.}~\bibnamefont {Guida}},\
  and\ \bibinfo {author} {\bibfnamefont {H.}~\bibnamefont {Umezawa}},\
  }\bibfield  {title} {\enquote {\bibinfo {title} {Dissipative quantum field
  theory - thermo field dynamics},}\ }\href@noop {} {\bibfield  {journal}
  {\bibinfo  {journal} {Physica A: Stat. Mech. Appl.}\ }\textbf {\bibinfo
  {volume} {148}},\ \bibinfo {pages} {1--26} (\bibinfo {year}
  {1988})}\BibitemShut {NoStop}%
\bibitem [{\citenamefont {Israel}(1976)}]{Israel.PLA.1976}%
  \BibitemOpen
  \bibfield  {author} {\bibinfo {author} {\bibfnamefont {W.}~\bibnamefont
  {Israel}},\ }\bibfield  {title} {\enquote {\bibinfo {title} {Thermo-field
  dynamics of black holes},}\ }\href@noop {} {\bibfield  {journal} {\bibinfo
  {journal} {Phys. Lett. A}\ }\textbf {\bibinfo {volume} {57}},\ \bibinfo
  {pages} {107--110} (\bibinfo {year} {1976})}\BibitemShut {NoStop}%
\bibitem [{\citenamefont {Laciana}(1994)}]{Laciana.GRG.1994}%
  \BibitemOpen
  \bibfield  {author} {\bibinfo {author} {\bibfnamefont {C.~E.}\ \bibnamefont
  {Laciana}},\ }\bibfield  {title} {\enquote {\bibinfo {title} {Quantum field
  theory in curved space-time as thermo field dynamics},}\ }\href@noop {}
  {\bibfield  {journal} {\bibinfo  {journal} {Gen Relat Gravit}\ }\textbf
  {\bibinfo {volume} {26}},\ \bibinfo {pages} {363--378} (\bibinfo {year}
  {1994})}\BibitemShut {NoStop}%
\bibitem [{\citenamefont {Matsumoto}\ and\ \citenamefont
  {Sakamoto}(2001)}]{Matsumoto.Sakamoto.PTP.2001}%
  \BibitemOpen
  \bibfield  {author} {\bibinfo {author} {\bibfnamefont {H.}~\bibnamefont
  {Matsumoto}}\ and\ \bibinfo {author} {\bibfnamefont {S.}~\bibnamefont
  {Sakamoto}},\ }\bibfield  {title} {\enquote {\bibinfo {title}
  {{Nonequilibrium Formulation in Bose-Einstein Condensed States}},}\
  }\href@noop {} {\bibfield  {journal} {\bibinfo  {journal} {Prog. Theor.
  Phys.}\ }\textbf {\bibinfo {volume} {105}},\ \bibinfo {pages} {573--590}
  (\bibinfo {year} {2001})}\BibitemShut {NoStop}%
\bibitem [{\citenamefont {da~Silva}\ \emph {et~al.}(2002)\citenamefont
  {da~Silva}, \citenamefont {Khanna}, \citenamefont {Matos~Neto},\ and\
  \citenamefont {Santana}}]{daSilva.etal.PhysRevA.2002}%
  \BibitemOpen
  \bibfield  {author} {\bibinfo {author} {\bibfnamefont {J.~C.}\ \bibnamefont
  {da~Silva}}, \bibinfo {author} {\bibfnamefont {F.~C.}\ \bibnamefont
  {Khanna}}, \bibinfo {author} {\bibfnamefont {A.}~\bibnamefont {Matos~Neto}},\
  and\ \bibinfo {author} {\bibfnamefont {A.~E.}\ \bibnamefont {Santana}},\
  }\bibfield  {title} {\enquote {\bibinfo {title} {Generalized bogoliubov
  transformation for confined fields: Applications for the casimir effect},}\
  }\href {https://doi.org/10.1103/PhysRevA.66.052101} {\bibfield  {journal}
  {\bibinfo  {journal} {Phys. Rev. A}\ }\textbf {\bibinfo {volume} {66}},\
  \bibinfo {pages} {052101} (\bibinfo {year} {2002})}\BibitemShut {NoStop}%
\bibitem [{\citenamefont {Chowdhury}(2013)}]{Chowdhury.IJMPD.2013}%
  \BibitemOpen
  \bibfield  {author} {\bibinfo {author} {\bibfnamefont {B.~D.}\ \bibnamefont
  {Chowdhury}},\ }\bibfield  {title} {\enquote {\bibinfo {title} {{Black Holes
  Versus Firewalls and Thermo-Field Dynamics}},}\ }\href@noop {} {\bibfield
  {journal} {\bibinfo  {journal} {Int. J. Mod. Phys. D}\ }\textbf {\bibinfo
  {volume} {22}},\ \bibinfo {pages} {1342011} (\bibinfo {year}
  {2013})}\BibitemShut {NoStop}%
\bibitem [{\citenamefont {Nair}(2015)}]{Nair.PhysRevD.2015}%
  \BibitemOpen
  \bibfield  {author} {\bibinfo {author} {\bibfnamefont {V.~P.}\ \bibnamefont
  {Nair}},\ }\bibfield  {title} {\enquote {\bibinfo {title} {Thermofield
  dynamics and gravity},}\ }\href {https://doi.org/10.1103/PhysRevD.92.104009}
  {\bibfield  {journal} {\bibinfo  {journal} {Phys. Rev. D}\ }\textbf {\bibinfo
  {volume} {92}},\ \bibinfo {pages} {104009} (\bibinfo {year}
  {2015})}\BibitemShut {NoStop}%
\bibitem [{\citenamefont {Yang}(2018)}]{Yang.PhysRevD.2018}%
  \BibitemOpen
  \bibfield  {author} {\bibinfo {author} {\bibfnamefont {R.-Q.}\ \bibnamefont
  {Yang}},\ }\bibfield  {title} {\enquote {\bibinfo {title} {Complexity for
  quantum field theory states and applications to thermofield double states},}\
  }\href {https://doi.org/10.1103/PhysRevD.97.066004} {\bibfield  {journal}
  {\bibinfo  {journal} {Phys. Rev. D}\ }\textbf {\bibinfo {volume} {97}},\
  \bibinfo {pages} {066004} (\bibinfo {year} {2018})}\BibitemShut {NoStop}%
\bibitem [{\citenamefont {Mu{\~n}oz~de Nova}\ \emph {et~al.}(2019)\citenamefont
  {Mu{\~n}oz~de Nova}, \citenamefont {Golubkov}, \citenamefont {Kolobov},\ and\
  \citenamefont {Steinhauer}}]{munoz2019observation}%
  \BibitemOpen
  \bibfield  {author} {\bibinfo {author} {\bibfnamefont {J.~R.}\ \bibnamefont
  {Mu{\~n}oz~de Nova}}, \bibinfo {author} {\bibfnamefont {K.}~\bibnamefont
  {Golubkov}}, \bibinfo {author} {\bibfnamefont {V.~I.}\ \bibnamefont
  {Kolobov}},\ and\ \bibinfo {author} {\bibfnamefont {J.}~\bibnamefont
  {Steinhauer}},\ }\bibfield  {title} {\enquote {\bibinfo {title} {Observation
  of thermal hawking radiation and its temperature in an analogue black
  hole},}\ }\href@noop {} {\bibfield  {journal} {\bibinfo  {journal} {Nature}\
  }\textbf {\bibinfo {volume} {569}},\ \bibinfo {pages} {688--691} (\bibinfo
  {year} {2019})}\BibitemShut {NoStop}%
\bibitem [{\citenamefont {Burrage}\ \emph {et~al.}(2019)\citenamefont
  {Burrage}, \citenamefont {K\"ading}, \citenamefont {Millington},\ and\
  \citenamefont {Min\'a\ifmmode~\check{r}\else
  \v{r}\fi{}}}]{Burrage.etal.PhysRevD.2019}%
  \BibitemOpen
  \bibfield  {author} {\bibinfo {author} {\bibfnamefont {C.}~\bibnamefont
  {Burrage}}, \bibinfo {author} {\bibfnamefont {C.}~\bibnamefont {K\"ading}},
  \bibinfo {author} {\bibfnamefont {P.}~\bibnamefont {Millington}},\ and\
  \bibinfo {author} {\bibfnamefont {J.~c.~v.}\ \bibnamefont
  {Min\'a\ifmmode~\check{r}\else \v{r}\fi{}}},\ }\bibfield  {title} {\enquote
  {\bibinfo {title} {Open quantum dynamics induced by light scalar fields},}\
  }\href {https://doi.org/10.1103/PhysRevD.100.076003} {\bibfield  {journal}
  {\bibinfo  {journal} {Phys. Rev. D}\ }\textbf {\bibinfo {volume} {100}},\
  \bibinfo {pages} {076003} (\bibinfo {year} {2019})}\BibitemShut {NoStop}%
\bibitem [{\citenamefont {Käding}\ and\ \citenamefont
  {Pitschmann}(2022)}]{Kading.Pitschmann.Universe.2022}%
  \BibitemOpen
  \bibfield  {author} {\bibinfo {author} {\bibfnamefont {C.}~\bibnamefont
  {Käding}}\ and\ \bibinfo {author} {\bibfnamefont {M.}~\bibnamefont
  {Pitschmann}},\ }\bibfield  {title} {\enquote {\bibinfo {title} {Density
  matrix formalism for interacting quantum fields},}\ }\href
  {https://doi.org/10.3390/universe8110601} {\bibfield  {journal} {\bibinfo
  {journal} {Universe}\ }\textbf {\bibinfo {volume} {8}} (\bibinfo {year}
  {2022}),\ 10.3390/universe8110601}\BibitemShut {NoStop}%
\bibitem [{\citenamefont {K\"ading}\ and\ \citenamefont
  {Pitschmann}(2023)}]{Kading.Pitschmann.PhysRevD.2023}%
  \BibitemOpen
  \bibfield  {author} {\bibinfo {author} {\bibfnamefont {C.}~\bibnamefont
  {K\"ading}}\ and\ \bibinfo {author} {\bibfnamefont {M.}~\bibnamefont
  {Pitschmann}},\ }\bibfield  {title} {\enquote {\bibinfo {title} {New method
  for directly computing reduced density matrices},}\ }\href
  {https://doi.org/10.1103/PhysRevD.107.016005} {\bibfield  {journal} {\bibinfo
   {journal} {Phys. Rev. D}\ }\textbf {\bibinfo {volume} {107}},\ \bibinfo
  {pages} {016005} (\bibinfo {year} {2023})}\BibitemShut {NoStop}%
\bibitem [{\citenamefont {Käding}\ and\ \citenamefont
  {Pitschmann}(2025)}]{Kading.Pitschmann.arXiv.2025}%
  \BibitemOpen
  \bibfield  {author} {\bibinfo {author} {\bibfnamefont {C.}~\bibnamefont
  {Käding}}\ and\ \bibinfo {author} {\bibfnamefont {M.}~\bibnamefont
  {Pitschmann}},\ }\href {https://arxiv.org/abs/2503.08567} {\enquote {\bibinfo
  {title} {Density matrices in quantum field theory: Non-markovianity, path
  integrals and master equations},}\ } (\bibinfo {year} {2025}),\ \Eprint
  {https://arxiv.org/abs/2503.08567} {arXiv:2503.08567 [hep-th]} \BibitemShut
  {NoStop}%
\bibitem [{\citenamefont {Chaturvedi}(1993)}]{Chaturvedi.1993}%
  \BibitemOpen
  \bibfield  {author} {\bibinfo {author} {\bibfnamefont {S.}~\bibnamefont
  {Chaturvedi}},\ }\bibfield  {title} {\enquote {\bibinfo {title} {Thermofield
  dynamics and its applications to quantum optics},}\ }in\ \href@noop {} {\emph
  {\bibinfo {booktitle} {Recent Developments in Quantum Optics}}}\ (\bibinfo
  {publisher} {Springer},\ \bibinfo {year} {1993})\ pp.\ \bibinfo {pages}
  {87--96}\BibitemShut {NoStop}%
\bibitem [{\citenamefont {Prud\^{e}ncio}(2012)}]{Prudencio.IJQI.2012}%
  \BibitemOpen
  \bibfield  {author} {\bibinfo {author} {\bibfnamefont {T.}~\bibnamefont
  {Prud\^{e}ncio}},\ }\bibfield  {title} {\enquote {\bibinfo {title}
  {No-{C}loning {T}heorem in {T}hermofield {D}ynamics},}\ }\href@noop {}
  {\bibfield  {journal} {\bibinfo  {journal} {Int. J. Quantum Inf.}\ }\textbf
  {\bibinfo {volume} {10}},\ \bibinfo {pages} {1230001} (\bibinfo {year}
  {2012})}\BibitemShut {NoStop}%
\bibitem [{\citenamefont {Wu}\ and\ \citenamefont
  {Hsieh}(2019)}]{Wu.Hsieh.PhysRevLett.2019}%
  \BibitemOpen
  \bibfield  {author} {\bibinfo {author} {\bibfnamefont {J.}~\bibnamefont
  {Wu}}\ and\ \bibinfo {author} {\bibfnamefont {T.~H.}\ \bibnamefont {Hsieh}},\
  }\bibfield  {title} {\enquote {\bibinfo {title} {Variational thermal quantum
  simulation via thermofield double states},}\ }\href
  {https://doi.org/10.1103/PhysRevLett.123.220502} {\bibfield  {journal}
  {\bibinfo  {journal} {Phys. Rev. Lett.}\ }\textbf {\bibinfo {volume} {123}},\
  \bibinfo {pages} {220502} (\bibinfo {year} {2019})}\BibitemShut {NoStop}%
\bibitem [{\citenamefont {del Campo}\ and\ \citenamefont
  {Takayanagi}(2020)}]{deCampo.Takayanagi.JHEP.2020}%
  \BibitemOpen
  \bibfield  {author} {\bibinfo {author} {\bibfnamefont {A.}~\bibnamefont {del
  Campo}}\ and\ \bibinfo {author} {\bibfnamefont {T.}~\bibnamefont
  {Takayanagi}},\ }\bibfield  {title} {\enquote {\bibinfo {title} {Decoherence
  in conformal field theory},}\ }\href@noop {} {\bibfield  {journal} {\bibinfo
  {journal} {J. High Energy Phys.}\ }\textbf {\bibinfo {volume} {2020}},\
  \bibinfo {pages} {1--27} (\bibinfo {year} {2020})}\BibitemShut {NoStop}%
\bibitem [{\citenamefont {Zhu}\ \emph {et~al.}(2020)\citenamefont {Zhu},
  \citenamefont {Johri}, \citenamefont {Linke}, \citenamefont {Landsman},
  \citenamefont {Alderete}, \citenamefont {Nguyen}, \citenamefont {Matsuura},
  \citenamefont {Hsieh},\ and\ \citenamefont {Monroe}}]{Zhu.etal.PNAS.2020}%
  \BibitemOpen
  \bibfield  {author} {\bibinfo {author} {\bibfnamefont {D.}~\bibnamefont
  {Zhu}}, \bibinfo {author} {\bibfnamefont {S.}~\bibnamefont {Johri}}, \bibinfo
  {author} {\bibfnamefont {N.~M.}\ \bibnamefont {Linke}}, \bibinfo {author}
  {\bibfnamefont {K.~A.}\ \bibnamefont {Landsman}}, \bibinfo {author}
  {\bibfnamefont {C.~H.}\ \bibnamefont {Alderete}}, \bibinfo {author}
  {\bibfnamefont {N.~H.}\ \bibnamefont {Nguyen}}, \bibinfo {author}
  {\bibfnamefont {A.~Y.}\ \bibnamefont {Matsuura}}, \bibinfo {author}
  {\bibfnamefont {T.~H.}\ \bibnamefont {Hsieh}},\ and\ \bibinfo {author}
  {\bibfnamefont {C.}~\bibnamefont {Monroe}},\ }\bibfield  {title} {\enquote
  {\bibinfo {title} {Generation of thermofield double states and critical
  ground states with a quantum computer},}\ }\href@noop {} {\bibfield
  {journal} {\bibinfo  {journal} {Proc. Natl. Acad. Sci. U.S.A.}\ }\textbf
  {\bibinfo {volume} {117}},\ \bibinfo {pages} {25402--25406} (\bibinfo {year}
  {2020})}\BibitemShut {NoStop}%
\bibitem [{\citenamefont {Xu}\ \emph {et~al.}(2021)\citenamefont {Xu},
  \citenamefont {Chenu}, \citenamefont {Prosen},\ and\ \citenamefont {del
  Campo}}]{Xu.etal.PRB.2021}%
  \BibitemOpen
  \bibfield  {author} {\bibinfo {author} {\bibfnamefont {Z.}~\bibnamefont
  {Xu}}, \bibinfo {author} {\bibfnamefont {A.}~\bibnamefont {Chenu}}, \bibinfo
  {author} {\bibfnamefont {T.~c.~v.}\ \bibnamefont {Prosen}},\ and\ \bibinfo
  {author} {\bibfnamefont {A.}~\bibnamefont {del Campo}},\ }\bibfield  {title}
  {\enquote {\bibinfo {title} {{Thermofield dynamics: Quantum chaos versus
  decoherence}},}\ }\href@noop {} {\bibfield  {journal} {\bibinfo  {journal}
  {Phys. Rev. B}\ }\textbf {\bibinfo {volume} {103}},\ \bibinfo {pages}
  {064309} (\bibinfo {year} {2021})}\BibitemShut {NoStop}%
\bibitem [{\citenamefont {Abidi}(2023)}]{Abidi.AmJPhysApp.2023}%
  \BibitemOpen
  \bibfield  {author} {\bibinfo {author} {\bibfnamefont {A.}~\bibnamefont
  {Abidi}},\ }\bibfield  {title} {\enquote {\bibinfo {title} {The thermo-field
  dynamics method for electron with two-mode electromagnetic field},}\ }\href
  {https://doi.org/10.11648/j.ajpa.20231101.13} {\bibfield  {journal} {\bibinfo
   {journal} {Am. J. Phys. Appl.}\ }\textbf {\bibinfo {volume} {11}},\ \bibinfo
  {pages} {21--30} (\bibinfo {year} {2023})},\ \Eprint
  {https://arxiv.org/abs/https://article.sciencepublishinggroup.com/pdf/10.11648.j.ajpa.20231101.13}
  {https://article.sciencepublishinggroup.com/pdf/10.11648.j.ajpa.20231101.13}
  \BibitemShut {NoStop}%
\bibitem [{\citenamefont {Petronilo}, \citenamefont {Araújo},\ and\
  \citenamefont {Cruz}(2023)}]{Petronilo.etal.2023}%
  \BibitemOpen
  \bibfield  {author} {\bibinfo {author} {\bibfnamefont {G.}~\bibnamefont
  {Petronilo}}, \bibinfo {author} {\bibfnamefont {M.}~\bibnamefont {Araújo}},\
  and\ \bibinfo {author} {\bibfnamefont {C.}~\bibnamefont {Cruz}},\ }\bibfield
  {title} {\enquote {\bibinfo {title} {Simulating thermal qubits through
  thermofield dynamics: an undergraduate approach using quantum computing},}\
  }\href {https://doi.org/10.1590/1806-9126-RBEF-2023-0287} {\bibfield
  {journal} {\bibinfo  {journal} {Rev Bras Ensino Fís}\ }\textbf {\bibinfo
  {volume} {45}},\ \bibinfo {pages} {e20230287} (\bibinfo {year}
  {2023})}\BibitemShut {NoStop}%
\bibitem [{\citenamefont {Nys}, \citenamefont {Denis},\ and\ \citenamefont
  {Carleo}(2024)}]{Nys.Denis.Carleo.PhysRevB.109.2024}%
  \BibitemOpen
  \bibfield  {author} {\bibinfo {author} {\bibfnamefont {J.}~\bibnamefont
  {Nys}}, \bibinfo {author} {\bibfnamefont {Z.}~\bibnamefont {Denis}},\ and\
  \bibinfo {author} {\bibfnamefont {G.}~\bibnamefont {Carleo}},\ }\bibfield
  {title} {\enquote {\bibinfo {title} {Real-time quantum dynamics of thermal
  states with neural thermofields},}\ }\href
  {https://doi.org/10.1103/PhysRevB.109.235120} {\bibfield  {journal} {\bibinfo
   {journal} {Phys. Rev. B}\ }\textbf {\bibinfo {volume} {109}},\ \bibinfo
  {pages} {235120} (\bibinfo {year} {2024})}\BibitemShut {NoStop}%
\bibitem [{\citenamefont {Liu}\ \emph {et~al.}(2024)\citenamefont {Liu},
  \citenamefont {Lv}, \citenamefont {Meng}, \citenamefont {Tan}, \citenamefont
  {Zhao},\ and\ \citenamefont {Zou}}]{Liu.etal.PhysRevResearch.2024}%
  \BibitemOpen
  \bibfield  {author} {\bibinfo {author} {\bibfnamefont {Y.}~\bibnamefont
  {Liu}}, \bibinfo {author} {\bibfnamefont {S.}~\bibnamefont {Lv}}, \bibinfo
  {author} {\bibfnamefont {Y.}~\bibnamefont {Meng}}, \bibinfo {author}
  {\bibfnamefont {Z.}~\bibnamefont {Tan}}, \bibinfo {author} {\bibfnamefont
  {E.}~\bibnamefont {Zhao}},\ and\ \bibinfo {author} {\bibfnamefont
  {H.}~\bibnamefont {Zou}},\ }\bibfield  {title} {\enquote {\bibinfo {title}
  {Exact fisher zeros and thermofield dynamics across a quantum critical
  point},}\ }\href {https://doi.org/10.1103/PhysRevResearch.6.043139}
  {\bibfield  {journal} {\bibinfo  {journal} {Phys. Rev. Res.}\ }\textbf
  {\bibinfo {volume} {6}},\ \bibinfo {pages} {043139} (\bibinfo {year}
  {2024})}\BibitemShut {NoStop}%
\bibitem [{\citenamefont {Harsha}, \citenamefont {Henderson},\ and\
  \citenamefont
  {Scuseria}(2019{\natexlab{a}})}]{Harsha.Henderson.Scuseria.JCP.2019}%
  \BibitemOpen
  \bibfield  {author} {\bibinfo {author} {\bibfnamefont {G.}~\bibnamefont
  {Harsha}}, \bibinfo {author} {\bibfnamefont {T.~M.}\ \bibnamefont
  {Henderson}},\ and\ \bibinfo {author} {\bibfnamefont {G.~E.}\ \bibnamefont
  {Scuseria}},\ }\bibfield  {title} {\enquote {\bibinfo {title} {Thermofield
  theory for finite-temperature quantum chemistry},}\ }\href
  {https://doi.org/10.1063/1.5089560} {\bibfield  {journal} {\bibinfo
  {journal} {J. Chem. Phys.}\ }\textbf {\bibinfo {volume} {150}},\ \bibinfo
  {pages} {154109} (\bibinfo {year} {2019}{\natexlab{a}})}\BibitemShut
  {NoStop}%
\bibitem [{\citenamefont {Harsha}, \citenamefont {Henderson},\ and\
  \citenamefont
  {Scuseria}(2019{\natexlab{b}})}]{Harsha.Henderson.Scuseria.JCTC.2019}%
  \BibitemOpen
  \bibfield  {author} {\bibinfo {author} {\bibfnamefont {G.}~\bibnamefont
  {Harsha}}, \bibinfo {author} {\bibfnamefont {T.~M.}\ \bibnamefont
  {Henderson}},\ and\ \bibinfo {author} {\bibfnamefont {G.~E.}\ \bibnamefont
  {Scuseria}},\ }\bibfield  {title} {\enquote {\bibinfo {title} {Thermofield
  theory for finite-temperature coupled cluster},}\ }\href@noop {} {\bibfield
  {journal} {\bibinfo  {journal} {J. Chem. Theory Comput.}\ }\textbf {\bibinfo
  {volume} {15}},\ \bibinfo {pages} {6127--6136} (\bibinfo {year}
  {2019}{\natexlab{b}})}\BibitemShut {NoStop}%
\bibitem [{\citenamefont {Bao}\ \emph {et~al.}(2024)\citenamefont {Bao},
  \citenamefont {Raymond}, \citenamefont {Zeng},\ and\ \citenamefont
  {Nooijen}}]{Bao.etal.JCTC.2024}%
  \BibitemOpen
  \bibfield  {author} {\bibinfo {author} {\bibfnamefont {S.}~\bibnamefont
  {Bao}}, \bibinfo {author} {\bibfnamefont {N.}~\bibnamefont {Raymond}},
  \bibinfo {author} {\bibfnamefont {T.}~\bibnamefont {Zeng}},\ and\ \bibinfo
  {author} {\bibfnamefont {M.}~\bibnamefont {Nooijen}},\ }\bibfield  {title}
  {\enquote {\bibinfo {title} {Vibrational electronic-thermofield coupled
  cluster (ve-tfcc) theory for quantum simulations of vibronic coupling systems
  at thermal equilibrium},}\ }\href@noop {} {\bibfield  {journal} {\bibinfo
  {journal} {J. Chem. Theory Comput.}\ }\textbf {\bibinfo {volume} {20}},\
  \bibinfo {pages} {5882--5900} (\bibinfo {year} {2024})}\BibitemShut {NoStop}%
\bibitem [{\citenamefont {Fischer}\ and\ \citenamefont
  {Saalfrank}(2021)}]{Fischer.Saalfrank.JCP.2021}%
  \BibitemOpen
  \bibfield  {author} {\bibinfo {author} {\bibfnamefont {E.~W.}\ \bibnamefont
  {Fischer}}\ and\ \bibinfo {author} {\bibfnamefont {P.}~\bibnamefont
  {Saalfrank}},\ }\bibfield  {title} {\enquote {\bibinfo {title} {{{A
  thermofield-based multilayer multiconfigurational time-dependent Hartree
  approach to non-adiabatic quantum dynamics at finite temperature}}},}\
  }\href@noop {} {\bibfield  {journal} {\bibinfo  {journal} {J. Chem. Phys.}\
  }\textbf {\bibinfo {volume} {155}},\ \bibinfo {pages} {134109} (\bibinfo
  {year} {2021})}\BibitemShut {NoStop}%
\bibitem [{\citenamefont {Borrelli}\ and\ \citenamefont
  {Gelin}(2021)}]{Borrelli.Gelin.WIRE.2021}%
  \BibitemOpen
  \bibfield  {author} {\bibinfo {author} {\bibfnamefont {R.}~\bibnamefont
  {Borrelli}}\ and\ \bibinfo {author} {\bibfnamefont {M.~F.}\ \bibnamefont
  {Gelin}},\ }\bibfield  {title} {\enquote {\bibinfo {title} {{Finite
  temperature quantum dynamics of complex systems: Integrating thermo-field
  theories and tensor-train methods}},}\ }\href@noop {} {\bibfield  {journal}
  {\bibinfo  {journal} {WIREs Comput. Mol. Sci.}\ }\textbf {\bibinfo {volume}
  {11}},\ \bibinfo {pages} {e1539} (\bibinfo {year} {2021})}\BibitemShut
  {NoStop}%
\bibitem [{\citenamefont {Chen}\ and\ \citenamefont
  {Zhao}(2017)}]{Chen.Zhao.JCP.2017}%
  \BibitemOpen
  \bibfield  {author} {\bibinfo {author} {\bibfnamefont {L.}~\bibnamefont
  {Chen}}\ and\ \bibinfo {author} {\bibfnamefont {Y.}~\bibnamefont {Zhao}},\
  }\bibfield  {title} {\enquote {\bibinfo {title} {{Finite temperature dynamics
  of a Holstein polaron: The thermo-field dynamics approach}},}\ }\href@noop {}
  {\bibfield  {journal} {\bibinfo  {journal} {J. Chem. Phys.}\ }\textbf
  {\bibinfo {volume} {147}},\ \bibinfo {pages} {214102} (\bibinfo {year}
  {2017})}\BibitemShut {NoStop}%
\bibitem [{\citenamefont {Borrelli}\ and\ \citenamefont
  {Gelin}(2017)}]{Borrelli.Gelin.SciRep.2017}%
  \BibitemOpen
  \bibfield  {author} {\bibinfo {author} {\bibfnamefont {R.}~\bibnamefont
  {Borrelli}}\ and\ \bibinfo {author} {\bibfnamefont {M.~F.}\ \bibnamefont
  {Gelin}},\ }\bibfield  {title} {\enquote {\bibinfo {title} {{Simulation of
  quantum dynamics of excitonic systems at finite temperature: An efficient
  method based on thermo field dynamics}},}\ }\href@noop {} {\bibfield
  {journal} {\bibinfo  {journal} {Sci. Rep.}\ }\textbf {\bibinfo {volume}
  {7}},\ \bibinfo {pages} {1--9} (\bibinfo {year} {2017})}\BibitemShut
  {NoStop}%
\bibitem [{\citenamefont {Brey}\ \emph {et~al.}(2021)\citenamefont {Brey},
  \citenamefont {Popp}, \citenamefont {Budakoti}, \citenamefont {D’Avino},\
  and\ \citenamefont {Burghardt}}]{Brey.etal.JPCC.2021}%
  \BibitemOpen
  \bibfield  {author} {\bibinfo {author} {\bibfnamefont {D.}~\bibnamefont
  {Brey}}, \bibinfo {author} {\bibfnamefont {W.}~\bibnamefont {Popp}}, \bibinfo
  {author} {\bibfnamefont {P.}~\bibnamefont {Budakoti}}, \bibinfo {author}
  {\bibfnamefont {G.}~\bibnamefont {D’Avino}},\ and\ \bibinfo {author}
  {\bibfnamefont {I.}~\bibnamefont {Burghardt}},\ }\bibfield  {title} {\enquote
  {\bibinfo {title} {{Quantum Dynamics of Electron–Hole Separation in Stacked
  Perylene Diimide-Based Self-Assembled Nanostructures}},}\ }\href@noop {}
  {\bibfield  {journal} {\bibinfo  {journal} {J. Phys. Chem. C}\ }\textbf
  {\bibinfo {volume} {125}},\ \bibinfo {pages} {25030--25043} (\bibinfo {year}
  {2021})}\BibitemShut {NoStop}%
\bibitem [{\citenamefont {Gelin}\ and\ \citenamefont
  {Borrelli}(2021)}]{Gelin.Borrelli.JCTC.2021}%
  \BibitemOpen
  \bibfield  {author} {\bibinfo {author} {\bibfnamefont {M.~F.}\ \bibnamefont
  {Gelin}}\ and\ \bibinfo {author} {\bibfnamefont {R.}~\bibnamefont
  {Borrelli}},\ }\bibfield  {title} {\enquote {\bibinfo {title} {{Simulation of
  Nonlinear Femtosecond Signals at Finite Temperature via a Thermo Field
  Dynamics-Tensor Train Method: General Theory and Application to Time- and
  Frequency-Resolved Fluorescence of the Fenna–Matthews–Olson Complex}},}\
  }\href@noop {} {\bibfield  {journal} {\bibinfo  {journal} {J. Chem. Theory
  Comput.}\ }\textbf {\bibinfo {volume} {17}},\ \bibinfo {pages} {4316--4331}
  (\bibinfo {year} {2021})}\BibitemShut {NoStop}%
\bibitem [{\citenamefont {Gelin}\ and\ \citenamefont
  {Borrelli}(2023)}]{Gelin.Borrelli.JCTC.2023}%
  \BibitemOpen
  \bibfield  {author} {\bibinfo {author} {\bibfnamefont {M.~F.}\ \bibnamefont
  {Gelin}}\ and\ \bibinfo {author} {\bibfnamefont {R.}~\bibnamefont
  {Borrelli}},\ }\bibfield  {title} {\enquote {\bibinfo {title} {{Thermo-Field
  Dynamics Approach to Photo-induced Electronic Transitions Driven by
  Incoherent Thermal Radiation}},}\ }\href@noop {} {\bibfield  {journal}
  {\bibinfo  {journal} {J. Chem. Theory Comput.}\ }\textbf {\bibinfo {volume}
  {19}},\ \bibinfo {pages} {6402--6413} (\bibinfo {year} {2023})}\BibitemShut
  {NoStop}%
\bibitem [{\citenamefont {Lyu}\ \emph {et~al.}(2024)\citenamefont {Lyu},
  \citenamefont {Khazaei}, \citenamefont {Geva},\ and\ \citenamefont
  {Batista}}]{Lyu.etal.JPCL.2024}%
  \BibitemOpen
  \bibfield  {author} {\bibinfo {author} {\bibfnamefont {N.}~\bibnamefont
  {Lyu}}, \bibinfo {author} {\bibfnamefont {P.}~\bibnamefont {Khazaei}},
  \bibinfo {author} {\bibfnamefont {E.}~\bibnamefont {Geva}},\ and\ \bibinfo
  {author} {\bibfnamefont {V.~S.}\ \bibnamefont {Batista}},\ }\bibfield
  {title} {\enquote {\bibinfo {title} {Simulating cavity-modified electron
  transfer dynamics on nisq computers},}\ }\href
  {https://doi.org/10.1021/acs.jpclett.4c02220} {\bibfield  {journal} {\bibinfo
   {journal} {J. Phys. Chem. Lett.}\ }\textbf {\bibinfo {volume} {15}},\
  \bibinfo {pages} {9535--9542} (\bibinfo {year} {2024})}\BibitemShut {NoStop}%
\bibitem [{\citenamefont {Chen}\ \emph {et~al.}(2021)\citenamefont {Chen},
  \citenamefont {Borrelli}, \citenamefont {Shalashilin}, \citenamefont {Zhao},\
  and\ \citenamefont {Gelin}}]{Chen.etal.JCTC.2021}%
  \BibitemOpen
  \bibfield  {author} {\bibinfo {author} {\bibfnamefont {L.}~\bibnamefont
  {Chen}}, \bibinfo {author} {\bibfnamefont {R.}~\bibnamefont {Borrelli}},
  \bibinfo {author} {\bibfnamefont {D.~V.}\ \bibnamefont {Shalashilin}},
  \bibinfo {author} {\bibfnamefont {Y.}~\bibnamefont {Zhao}},\ and\ \bibinfo
  {author} {\bibfnamefont {M.~F.}\ \bibnamefont {Gelin}},\ }\bibfield  {title}
  {\enquote {\bibinfo {title} {{Simulation of Time- and Frequency-Resolved
  Four-Wave-Mixing Signals at Finite Temperatures: A Thermo-Field Dynamics
  Approach}},}\ }\href@noop {} {\bibfield  {journal} {\bibinfo  {journal} {J.
  Chem. Theory Comput.}\ }\textbf {\bibinfo {volume} {17}},\ \bibinfo {pages}
  {4359--4373} (\bibinfo {year} {2021})}\BibitemShut {NoStop}%
\bibitem [{\citenamefont {Begu\v{s}i{\'c}}\ and\ \citenamefont
  {Van{\'i}\v{c}ek}(2021)}]{Begusic.Vanicek.JPCL.2021}%
  \BibitemOpen
  \bibfield  {author} {\bibinfo {author} {\bibfnamefont {T.}~\bibnamefont
  {Begu\v{s}i{\'c}}}\ and\ \bibinfo {author} {\bibfnamefont {J.}~\bibnamefont
  {Van{\'i}\v{c}ek}},\ }\bibfield  {title} {\enquote {\bibinfo {title}
  {{Finite-Temperature, Anharmonicity, and Duschinsky Effects on the
  Two-Dimensional Electronic Spectra from Ab Initio Thermo-Field Gaussian
  Wavepacket Dynamics}},}\ }\href@noop {} {\bibfield  {journal} {\bibinfo
  {journal} {J. Phys. Chem. Lett.}\ }\textbf {\bibinfo {volume} {12}},\
  \bibinfo {pages} {2997--3005} (\bibinfo {year} {2021})}\BibitemShut {NoStop}%
\bibitem [{\citenamefont {Zhang}\ and\ \citenamefont {Van{\'
  i}\v{c}ek}(2024)}]{Zhang.Vanicek.JCP.2024}%
  \BibitemOpen
  \bibfield  {author} {\bibinfo {author} {\bibfnamefont {Z.~T.}\ \bibnamefont
  {Zhang}}\ and\ \bibinfo {author} {\bibfnamefont {J.~J.~L.}\ \bibnamefont
  {Van{\' i}\v{c}ek}},\ }\bibfield  {title} {\enquote {\bibinfo {title}
  {{Finite-temperature vibronic spectra from the split-operator coherence
  thermofield dynamics}},}\ }\href@noop {} {\bibfield  {journal} {\bibinfo
  {journal} {J. Chem. Phys.}\ }\textbf {\bibinfo {volume} {160}},\ \bibinfo
  {pages} {084103} (\bibinfo {year} {2024})}\BibitemShut {NoStop}%
\bibitem [{\citenamefont {Polley}\ and\ \citenamefont
  {Loring}(2022)}]{Polley.Loring.JCP.2022}%
  \BibitemOpen
  \bibfield  {author} {\bibinfo {author} {\bibfnamefont {K.}~\bibnamefont
  {Polley}}\ and\ \bibinfo {author} {\bibfnamefont {R.~F.}\ \bibnamefont
  {Loring}},\ }\bibfield  {title} {\enquote {\bibinfo {title} {{Two-dimensional
  vibronic spectroscopy with semiclassical thermofield dynamics}},}\
  }\href@noop {} {\bibfield  {journal} {\bibinfo  {journal} {J. Chem. Phys.}\
  }\textbf {\bibinfo {volume} {156}},\ \bibinfo {pages} {124108} (\bibinfo
  {year} {2022})}\BibitemShut {NoStop}%
\bibitem [{\citenamefont {Barnett}\ and\ \citenamefont
  {Knight}(1985)}]{Barnett85}%
  \BibitemOpen
  \bibfield  {author} {\bibinfo {author} {\bibfnamefont {S.~M.}\ \bibnamefont
  {Barnett}}\ and\ \bibinfo {author} {\bibfnamefont {P.~L.}\ \bibnamefont
  {Knight}},\ }\bibfield  {title} {\enquote {\bibinfo {title} {Thermofield
  analysis of squeezing and statistical mixtures in quantum optics},}\
  }\href@noop {} {\bibfield  {journal} {\bibinfo  {journal} {J. Opt. Soc. Am.
  B}\ }\textbf {\bibinfo {volume} {2}},\ \bibinfo {pages} {467} (\bibinfo
  {year} {1985})}\BibitemShut {NoStop}%
\bibitem [{\citenamefont {Gelin}\ and\ \citenamefont
  {Borrelli}(2017)}]{Gelin.Borrelli.AnnPhys.2017}%
  \BibitemOpen
  \bibfield  {author} {\bibinfo {author} {\bibfnamefont {M.~F.}\ \bibnamefont
  {Gelin}}\ and\ \bibinfo {author} {\bibfnamefont {R.}~\bibnamefont
  {Borrelli}},\ }\bibfield  {title} {\enquote {\bibinfo {title} {{Thermal
  Schrödinger Equation: Efficient Tool for Simulation of Many-Body Quantum
  Dynamics at Finite Temperature}},}\ }\href@noop {} {\bibfield  {journal}
  {\bibinfo  {journal} {Ann. Phys.}\ }\textbf {\bibinfo {volume} {529}},\
  \bibinfo {pages} {1700200} (\bibinfo {year} {2017})}\BibitemShut {NoStop}%
\bibitem [{\citenamefont {Błasiak}\ \emph
  {et~al.}(2025{\natexlab{a}})\citenamefont {Błasiak}, \citenamefont {Brey},
  \citenamefont {Martinazzo},\ and\ \citenamefont {Burghardt}}]{BBMB-1-2025}%
  \BibitemOpen
  \bibfield  {author} {\bibinfo {author} {\bibfnamefont {B.}~\bibnamefont
  {Błasiak}}, \bibinfo {author} {\bibfnamefont {D.}~\bibnamefont {Brey}},
  \bibinfo {author} {\bibfnamefont {R.}~\bibnamefont {Martinazzo}},\ and\
  \bibinfo {author} {\bibfnamefont {I.}~\bibnamefont {Burghardt}},\ }\bibfield
  {title} {\enquote {\bibinfo {title} {Quantum dynamics at conical
  intersections in solution. i. multiplicative neural networks and
  thermofields},}\ }\href {https://doi.org/10.1063/5.0284503} {\bibfield
  {journal} {\bibinfo  {journal} {The Journal of Chemical Physics}\ }\textbf
  {\bibinfo {volume} {163}},\ \bibinfo {pages} {124108} (\bibinfo {year}
  {2025}{\natexlab{a}})}\BibitemShut {NoStop}%
\bibitem [{\citenamefont {Błasiak}\ \emph
  {et~al.}(2025{\natexlab{b}})\citenamefont {Błasiak}, \citenamefont {Brey},
  \citenamefont {Martinazzo},\ and\ \citenamefont {Burghardt}}]{BBMB-2-2025}%
  \BibitemOpen
  \bibfield  {author} {\bibinfo {author} {\bibfnamefont {B.}~\bibnamefont
  {Błasiak}}, \bibinfo {author} {\bibfnamefont {D.}~\bibnamefont {Brey}},
  \bibinfo {author} {\bibfnamefont {R.}~\bibnamefont {Martinazzo}},\ and\
  \bibinfo {author} {\bibfnamefont {I.}~\bibnamefont {Burghardt}},\ }\bibfield
  {title} {\enquote {\bibinfo {title} {Quantum dynamics at conical
  intersections in solution. ii. multiconfigurational wavefunction dynamics at
  finite temperature},}\ }\href {https://doi.org/10.1063/5.0284504} {\bibfield
  {journal} {\bibinfo  {journal} {The Journal of Chemical Physics}\ }\textbf
  {\bibinfo {volume} {163}},\ \bibinfo {pages} {124109} (\bibinfo {year}
  {2025}{\natexlab{b}})}\BibitemShut {NoStop}%
\bibitem [{\citenamefont {Clerk}\ \emph {et~al.}(2010)\citenamefont {Clerk},
  \citenamefont {Devoret}, \citenamefont {Girvin}, \citenamefont {Marquardt},\
  and\ \citenamefont {Schoelkopf}}]{Clerk2010}%
  \BibitemOpen
  \bibfield  {author} {\bibinfo {author} {\bibfnamefont {A.~A.}\ \bibnamefont
  {Clerk}}, \bibinfo {author} {\bibfnamefont {M.~H.}\ \bibnamefont {Devoret}},
  \bibinfo {author} {\bibfnamefont {S.~M.}\ \bibnamefont {Girvin}}, \bibinfo
  {author} {\bibfnamefont {F.}~\bibnamefont {Marquardt}},\ and\ \bibinfo
  {author} {\bibfnamefont {R.~J.}\ \bibnamefont {Schoelkopf}},\ }\bibfield
  {title} {\enquote {\bibinfo {title} {Introduction to quantum noise,
  measurement, and amplification},}\ }\href
  {https://doi.org/10.1103/RevModPhys.82.1155} {\bibfield  {journal} {\bibinfo
  {journal} {Rev. Mod. Phys.}\ }\textbf {\bibinfo {volume} {82}},\ \bibinfo
  {pages} {1155--1208} (\bibinfo {year} {2010})}\BibitemShut {NoStop}%
\bibitem [{\citenamefont {Tamascelli}\ \emph {et~al.}(2019)\citenamefont
  {Tamascelli}, \citenamefont {Smirne}, \citenamefont {Lim}, \citenamefont
  {Huelga},\ and\ \citenamefont {Plenio}}]{Tamascelli2019}%
  \BibitemOpen
  \bibfield  {author} {\bibinfo {author} {\bibfnamefont {D.}~\bibnamefont
  {Tamascelli}}, \bibinfo {author} {\bibfnamefont {A.}~\bibnamefont {Smirne}},
  \bibinfo {author} {\bibfnamefont {J.}~\bibnamefont {Lim}}, \bibinfo {author}
  {\bibfnamefont {S.}~\bibnamefont {Huelga}},\ and\ \bibinfo {author}
  {\bibfnamefont {M.}~\bibnamefont {Plenio}},\ }\bibfield  {title} {\enquote
  {\bibinfo {title} {Efficient simulation of finite-temperature open quantum
  systems},}\ }\href {https://doi.org/10.1103/physrevlett.123.090402}
  {\bibfield  {journal} {\bibinfo  {journal} {Phys. Rev. Lett.}\ }\textbf
  {\bibinfo {volume} {123}},\ \bibinfo {pages} {090402} (\bibinfo {year}
  {2019})}\BibitemShut {NoStop}%
\bibitem [{\citenamefont {Takahashi}\ and\ \citenamefont
  {Borrelli}(2024)}]{Takahashi2024}%
  \BibitemOpen
  \bibfield  {author} {\bibinfo {author} {\bibfnamefont {H.}~\bibnamefont
  {Takahashi}}\ and\ \bibinfo {author} {\bibfnamefont {R.}~\bibnamefont
  {Borrelli}},\ }\bibfield  {title} {\enquote {\bibinfo {title} {Effective
  modeling of open quantum systems by low-rank discretization of structured
  environments},}\ }\href {https://doi.org/10.1063/5.0232232} {\bibfield
  {journal} {\bibinfo  {journal} {J. Chem. Phys.}\ }\textbf {\bibinfo {volume}
  {161}},\ \bibinfo {pages} {151101} (\bibinfo {year} {2024})}\BibitemShut
  {NoStop}%
\bibitem [{\citenamefont {Matsumoto}, \citenamefont {Nakano},\ and\
  \citenamefont {Umezawa}(1985)}]{Matsumoto.Nakano.Umezawa.PRD.1985}%
  \BibitemOpen
  \bibfield  {author} {\bibinfo {author} {\bibfnamefont {H.}~\bibnamefont
  {Matsumoto}}, \bibinfo {author} {\bibfnamefont {Y.}~\bibnamefont {Nakano}},\
  and\ \bibinfo {author} {\bibfnamefont {H.}~\bibnamefont {Umezawa}},\
  }\bibfield  {title} {\enquote {\bibinfo {title} {{Tilde substitution law in
  thermo field dynamics: Thermal state conditions}},}\ }\href@noop {}
  {\bibfield  {journal} {\bibinfo  {journal} {Phys. Rev. D}\ }\textbf {\bibinfo
  {volume} {31}},\ \bibinfo {pages} {429--432} (\bibinfo {year}
  {1985})}\BibitemShut {NoStop}%
\bibitem [{\citenamefont {Mann}\ \emph {et~al.}(1989)\citenamefont {Mann},
  \citenamefont {Revzen}, \citenamefont {Nakamura}, \citenamefont {Umezawa},\
  and\ \citenamefont {Yamanaka}}]{mann1989coherent}%
  \BibitemOpen
  \bibfield  {author} {\bibinfo {author} {\bibfnamefont {A.}~\bibnamefont
  {Mann}}, \bibinfo {author} {\bibfnamefont {M.}~\bibnamefont {Revzen}},
  \bibinfo {author} {\bibfnamefont {K.}~\bibnamefont {Nakamura}}, \bibinfo
  {author} {\bibfnamefont {H.}~\bibnamefont {Umezawa}},\ and\ \bibinfo {author}
  {\bibfnamefont {Y.}~\bibnamefont {Yamanaka}},\ }\bibfield  {title} {\enquote
  {\bibinfo {title} {Coherent and thermal coherent state},}\ }\href@noop {}
  {\bibfield  {journal} {\bibinfo  {journal} {J. Math. Phys.}\ }\textbf
  {\bibinfo {volume} {30}},\ \bibinfo {pages} {2883--2890} (\bibinfo {year}
  {1989})}\BibitemShut {NoStop}%
\bibitem [{\citenamefont {Beck}\ \emph {et~al.}(2000)\citenamefont {Beck},
  \citenamefont {J{\"a}ckle}, \citenamefont {Worth},\ and\ \citenamefont
  {Meyer}}]{Beck2000}%
  \BibitemOpen
  \bibfield  {author} {\bibinfo {author} {\bibfnamefont {M.~H.}\ \bibnamefont
  {Beck}}, \bibinfo {author} {\bibfnamefont {A.}~\bibnamefont {J{\"a}ckle}},
  \bibinfo {author} {\bibfnamefont {G.~A.}\ \bibnamefont {Worth}},\ and\
  \bibinfo {author} {\bibfnamefont {H.-D.}\ \bibnamefont {Meyer}},\ }\bibfield
  {title} {\enquote {\bibinfo {title} {The multiconfiguration time-dependent
  hartree (mctdh) method: a highly efficient algorithm for propagating
  wavepackets},}\ }\href@noop {} {\bibfield  {journal} {\bibinfo  {journal}
  {Phys. Rep.}\ }\textbf {\bibinfo {volume} {324}},\ \bibinfo {pages} {1}
  (\bibinfo {year} {2000})}\BibitemShut {NoStop}%
\bibitem [{\citenamefont {Wang}(2015)}]{Wang2015}%
  \BibitemOpen
  \bibfield  {author} {\bibinfo {author} {\bibfnamefont {H.}~\bibnamefont
  {Wang}},\ }\bibfield  {title} {\enquote {\bibinfo {title} {Multilayer
  multiconfiguration time-dependent hartree theory},}\ }\href@noop {}
  {\bibfield  {journal} {\bibinfo  {journal} {J. Phys. Chem. A}\ }\textbf
  {\bibinfo {volume} {119}},\ \bibinfo {pages} {7951} (\bibinfo {year}
  {2015})}\BibitemShut {NoStop}%
\bibitem [{\citenamefont {Elmfors}\ and\ \citenamefont
  {Umezawa}(1994)}]{Elmfors.Umezawa.PA.1994}%
  \BibitemOpen
  \bibfield  {author} {\bibinfo {author} {\bibfnamefont {P.}~\bibnamefont
  {Elmfors}}\ and\ \bibinfo {author} {\bibfnamefont {H.}~\bibnamefont
  {Umezawa}},\ }\bibfield  {title} {\enquote {\bibinfo {title}
  {{Generalizations of the thermal Bogoliubov transformation}},}\ }\href@noop
  {} {\bibfield  {journal} {\bibinfo  {journal} {Physica A: Stat. Mech. Appl.}\
  }\textbf {\bibinfo {volume} {202}},\ \bibinfo {pages} {577--594} (\bibinfo
  {year} {1994})}\BibitemShut {NoStop}%
\bibitem [{\citenamefont {Caldeira}\ and\ \citenamefont
  {Leggett}(1983)}]{Caldeira.Leggett.AnnPhys.1983}%
  \BibitemOpen
  \bibfield  {author} {\bibinfo {author} {\bibfnamefont {A.}~\bibnamefont
  {Caldeira}}\ and\ \bibinfo {author} {\bibfnamefont {A.}~\bibnamefont
  {Leggett}},\ }\bibfield  {title} {\enquote {\bibinfo {title} {Quantum
  tunnelling in a dissipative system},}\ }\href@noop {} {\bibfield  {journal}
  {\bibinfo  {journal} {Ann. Phys.}\ }\textbf {\bibinfo {volume} {149}},\
  \bibinfo {pages} {374--456} (\bibinfo {year} {1983})}\BibitemShut {NoStop}%
\bibitem [{\citenamefont {Satyanarayana}(1985)}]{Satyanarayana.PRD.1985}%
  \BibitemOpen
  \bibfield  {author} {\bibinfo {author} {\bibfnamefont {M.~V.}\ \bibnamefont
  {Satyanarayana}},\ }\bibfield  {title} {\enquote {\bibinfo {title}
  {Generalized coherent states and generalized squeezed coherent states},}\
  }\href@noop {} {\bibfield  {journal} {\bibinfo  {journal} {Phys. Rev. D}\
  }\textbf {\bibinfo {volume} {32}},\ \bibinfo {pages} {400--404} (\bibinfo
  {year} {1985})}\BibitemShut {NoStop}%
\bibitem [{\citenamefont {Kim}, \citenamefont {de~Oliveira},\ and\
  \citenamefont {Knight}(1989)}]{Kim.etal.PRA.1989}%
  \BibitemOpen
  \bibfield  {author} {\bibinfo {author} {\bibfnamefont {M.~S.}\ \bibnamefont
  {Kim}}, \bibinfo {author} {\bibfnamefont {F.~A.~M.}\ \bibnamefont
  {de~Oliveira}},\ and\ \bibinfo {author} {\bibfnamefont {P.~L.}\ \bibnamefont
  {Knight}},\ }\bibfield  {title} {\enquote {\bibinfo {title} {Properties of
  squeezed number states and squeezed thermal states},}\ }\href@noop {}
  {\bibfield  {journal} {\bibinfo  {journal} {Phys. Rev. A}\ }\textbf {\bibinfo
  {volume} {40}},\ \bibinfo {pages} {2494--2503} (\bibinfo {year}
  {1989})}\BibitemShut {NoStop}%
\bibitem [{\citenamefont {de~Oliveira}\ \emph {et~al.}(1990)\citenamefont
  {de~Oliveira}, \citenamefont {Kim}, \citenamefont {Knight},\ and\
  \citenamefont {Bu\ifmmode~\check{z}\else
  \v{z}\fi{}ek}}]{Oliviera.etal.PRA.1990}%
  \BibitemOpen
  \bibfield  {author} {\bibinfo {author} {\bibfnamefont {F.~A.~M.}\
  \bibnamefont {de~Oliveira}}, \bibinfo {author} {\bibfnamefont {M.~S.}\
  \bibnamefont {Kim}}, \bibinfo {author} {\bibfnamefont {P.~L.}\ \bibnamefont
  {Knight}},\ and\ \bibinfo {author} {\bibfnamefont {V.}~\bibnamefont
  {Bu\ifmmode~\check{z}\else \v{z}\fi{}ek}},\ }\bibfield  {title} {\enquote
  {\bibinfo {title} {Properties of displaced number states},}\ }\href@noop {}
  {\bibfield  {journal} {\bibinfo  {journal} {Phys. Rev. A}\ }\textbf {\bibinfo
  {volume} {41}},\ \bibinfo {pages} {2645--2652} (\bibinfo {year}
  {1990})}\BibitemShut {NoStop}%
\bibitem [{\citenamefont {Zhang}, \citenamefont {Gilmore}\ \emph
  {et~al.}(1990)\citenamefont {Zhang}, \citenamefont {Gilmore} \emph
  {et~al.}}]{zhang1990coherent}%
  \BibitemOpen
  \bibfield  {author} {\bibinfo {author} {\bibfnamefont {W.-M.}\ \bibnamefont
  {Zhang}}, \bibinfo {author} {\bibfnamefont {R.}~\bibnamefont {Gilmore}},
  \emph {et~al.},\ }\bibfield  {title} {\enquote {\bibinfo {title} {{Coherent
  states: Theory and some applications}},}\ }\href@noop {} {\bibfield
  {journal} {\bibinfo  {journal} {Rev. Mod. Phys.}\ }\textbf {\bibinfo {volume}
  {62}},\ \bibinfo {pages} {867} (\bibinfo {year} {1990})}\BibitemShut
  {NoStop}%
\bibitem [{\citenamefont {Malbouisson}, \citenamefont {Baseia},\ and\
  \citenamefont {Avelar}(2007)}]{Malbouisson.Baseia.Avelar.PhysA.2007}%
  \BibitemOpen
  \bibfield  {author} {\bibinfo {author} {\bibfnamefont {J.}~\bibnamefont
  {Malbouisson}}, \bibinfo {author} {\bibfnamefont {B.}~\bibnamefont
  {Baseia}},\ and\ \bibinfo {author} {\bibfnamefont {A.}~\bibnamefont
  {Avelar}},\ }\bibfield  {title} {\enquote {\bibinfo {title} {A note on the
  generation of displaced number states},}\ }\href@noop {} {\bibfield
  {journal} {\bibinfo  {journal} {Physica A: Stat. Mech. Appl.}\ }\textbf
  {\bibinfo {volume} {376}},\ \bibinfo {pages} {275--278} (\bibinfo {year}
  {2007})}\BibitemShut {NoStop}%
\bibitem [{\citenamefont {Hsu}(2020)}]{Hsu.PCCP.2020}%
  \BibitemOpen
  \bibfield  {author} {\bibinfo {author} {\bibfnamefont {C.-P.}\ \bibnamefont
  {Hsu}},\ }\bibfield  {title} {\enquote {\bibinfo {title} {Reorganization
  energies and spectral densities for electron transfer problems in charge
  transport materials},}\ }\href@noop {} {\bibfield  {journal} {\bibinfo
  {journal} {Phys. Chem. Chem. Phys.}\ }\textbf {\bibinfo {volume} {22}},\
  \bibinfo {pages} {21630--21641} (\bibinfo {year} {2020})}\BibitemShut
  {NoStop}%
\bibitem [{\citenamefont {Butkus}, \citenamefont {Valkunas},\ and\
  \citenamefont {Abramavicius}(2012)}]{Butkus.Valkunas.Abramavicius.JCP.2012}%
  \BibitemOpen
  \bibfield  {author} {\bibinfo {author} {\bibfnamefont {V.}~\bibnamefont
  {Butkus}}, \bibinfo {author} {\bibfnamefont {L.}~\bibnamefont {Valkunas}},\
  and\ \bibinfo {author} {\bibfnamefont {D.}~\bibnamefont {Abramavicius}},\
  }\bibfield  {title} {\enquote {\bibinfo {title} {{Molecular
  vibrations-induced quantum beats in two-dimensional electronic
  spectroscopy}},}\ }\href@noop {} {\bibfield  {journal} {\bibinfo  {journal}
  {J. Chem. Phys.}\ }\textbf {\bibinfo {volume} {137}},\ \bibinfo {pages}
  {044513} (\bibinfo {year} {2012})}\BibitemShut {NoStop}%
\bibitem [{\citenamefont {V{\"o}hringer}\ \emph {et~al.}(1995)\citenamefont
  {V{\"o}hringer}, \citenamefont {Arnett}, \citenamefont {Westervelt},
  \citenamefont {Feldstein},\ and\ \citenamefont
  {Scherer}}]{Voerhinger.etal.JCP.1995}%
  \BibitemOpen
  \bibfield  {author} {\bibinfo {author} {\bibfnamefont {P.}~\bibnamefont
  {V{\"o}hringer}}, \bibinfo {author} {\bibfnamefont {D.~C.}\ \bibnamefont
  {Arnett}}, \bibinfo {author} {\bibfnamefont {R.~A.}\ \bibnamefont
  {Westervelt}}, \bibinfo {author} {\bibfnamefont {M.~J.}\ \bibnamefont
  {Feldstein}},\ and\ \bibinfo {author} {\bibfnamefont {N.~F.}\ \bibnamefont
  {Scherer}},\ }\bibfield  {title} {\enquote {\bibinfo {title} {{Optical
  dephasing on femtosecond time scales: Direct measurement and calculation from
  solvent spectral densities}},}\ }\href@noop {} {\bibfield  {journal}
  {\bibinfo  {journal} {J. Chem. Phys.}\ }\textbf {\bibinfo {volume} {102}},\
  \bibinfo {pages} {4027--4036} (\bibinfo {year} {1995})}\BibitemShut {NoStop}%
\bibitem [{\citenamefont {Cho}\ \emph {et~al.}(1993)\citenamefont {Cho},
  \citenamefont {Du}, \citenamefont {Scherer}, \citenamefont {Fleming},\ and\
  \citenamefont {Mukamel}}]{Cho.etal.JCP.1993}%
  \BibitemOpen
  \bibfield  {author} {\bibinfo {author} {\bibfnamefont {M.}~\bibnamefont
  {Cho}}, \bibinfo {author} {\bibfnamefont {M.}~\bibnamefont {Du}}, \bibinfo
  {author} {\bibfnamefont {N.~F.}\ \bibnamefont {Scherer}}, \bibinfo {author}
  {\bibfnamefont {G.~R.}\ \bibnamefont {Fleming}},\ and\ \bibinfo {author}
  {\bibfnamefont {S.}~\bibnamefont {Mukamel}},\ }\bibfield  {title} {\enquote
  {\bibinfo {title} {{Off‐resonant transient birefringence in liquids}},}\
  }\href@noop {} {\bibfield  {journal} {\bibinfo  {journal} {J. Chem. Phys.}\
  }\textbf {\bibinfo {volume} {99}},\ \bibinfo {pages} {2410--2428} (\bibinfo
  {year} {1993})}\BibitemShut {NoStop}%
\bibitem [{\citenamefont {Lu}(1999)}]{W-F-Lu-1999}%
  \BibitemOpen
  \bibfield  {author} {\bibinfo {author} {\bibfnamefont {W.-F.}\ \bibnamefont
  {Lu}},\ }\bibfield  {title} {\enquote {\bibinfo {title} {Thermalized
  displaced and squeezed number states in the coordinate representation},}\
  }\href {https://doi.org/10.1088/0305-4470/32/27/305} {\bibfield  {journal}
  {\bibinfo  {journal} {J. Phys. A: Math. Gen.}\ }\textbf {\bibinfo {volume}
  {32}},\ \bibinfo {pages} {5037} (\bibinfo {year} {1999})}\BibitemShut
  {NoStop}%
\bibitem [{\citenamefont {Tay}(2011)}]{Tay2011}%
  \BibitemOpen
  \bibfield  {author} {\bibinfo {author} {\bibfnamefont {B.~A.}\ \bibnamefont
  {Tay}},\ }\bibfield  {title} {\enquote {\bibinfo {title} {{Transformed number
  states from a generalized Bogoliubov transformation and their relationships
  to the eigenstates of the Kossakowski–Lindblad equation}},}\ }\href@noop {}
  {\bibfield  {journal} {\bibinfo  {journal} {J. Phys. A: Math. Theor.}\
  }\textbf {\bibinfo {volume} {44}},\ \bibinfo {pages} {255303} (\bibinfo
  {year} {2011})}\BibitemShut {NoStop}%
\bibitem [{{\relax DLMF}()}]{NIST:DLMF}%
  \BibitemOpen
  {\relax DLMF},\ \href {https://dlmf.nist.gov/} {\enquote {\bibinfo {title}
  {{\it NIST Digital Library of Mathematical Functions}},}\ }\bibinfo
  {howpublished} {\url{https://dlmf.nist.gov/}, Release 1.2.1 of 2024-06-15},\
  \bibinfo {note} {f.~W.~J. Olver, A.~B. {Olde Daalhuis}, D.~W. Lozier, B.~I.
  Schneider, R.~F. Boisvert, C.~W. Clark, B.~R. Miller, B.~V. Saunders, H.~S.
  Cohl, and M.~A. McClain, eds.}\BibitemShut {Stop}%
\bibitem [{\citenamefont {Hughes}, \citenamefont {Parry},\ and\ \citenamefont
  {Burghardt}(2009)}]{Hughes.etal.JCP.2009}%
  \BibitemOpen
  \bibfield  {author} {\bibinfo {author} {\bibfnamefont {K.~H.}\ \bibnamefont
  {Hughes}}, \bibinfo {author} {\bibfnamefont {S.~M.}\ \bibnamefont {Parry}},\
  and\ \bibinfo {author} {\bibfnamefont {I.}~\bibnamefont {Burghardt}},\
  }\bibfield  {title} {\enquote {\bibinfo {title} {{Closure of quantum
  hydrodynamic moment equations}},}\ }\href {https://doi.org/10.1063/1.3073759}
  {\bibfield  {journal} {\bibinfo  {journal} {J. Chem. Phys.}\ }\textbf
  {\bibinfo {volume} {130}},\ \bibinfo {pages} {054115} (\bibinfo {year}
  {2009})}\BibitemShut {NoStop}%
\bibitem [{\citenamefont {Caves}\ and\ \citenamefont
  {Schumaker}(1985)}]{Caves.Schumaker.PRA.1985}%
  \BibitemOpen
  \bibfield  {author} {\bibinfo {author} {\bibfnamefont {C.~M.}\ \bibnamefont
  {Caves}}\ and\ \bibinfo {author} {\bibfnamefont {B.~L.}\ \bibnamefont
  {Schumaker}},\ }\bibfield  {title} {\enquote {\bibinfo {title} {{New
  formalism for two-photon quantum optics. I. Quadrature phases and squeezed
  states}},}\ }\href {https://doi.org/10.1103/PhysRevA.31.3068} {\bibfield
  {journal} {\bibinfo  {journal} {Phys. Rev. A}\ }\textbf {\bibinfo {volume}
  {31}},\ \bibinfo {pages} {3068--3092} (\bibinfo {year} {1985})}\BibitemShut
  {NoStop}%
\bibitem [{\citenamefont {Wang}\ \emph {et~al.}(2024)\citenamefont {Wang},
  \citenamefont {Zhan}, \citenamefont {Vetlugin}, \citenamefont {Ou},
  \citenamefont {Liu}, \citenamefont {Shen},\ and\ \citenamefont
  {Fu}}]{wang2024structured}%
  \BibitemOpen
  \bibfield  {author} {\bibinfo {author} {\bibfnamefont {Z.}~\bibnamefont
  {Wang}}, \bibinfo {author} {\bibfnamefont {Z.}~\bibnamefont {Zhan}}, \bibinfo
  {author} {\bibfnamefont {A.~N.}\ \bibnamefont {Vetlugin}}, \bibinfo {author}
  {\bibfnamefont {J.-Y.}\ \bibnamefont {Ou}}, \bibinfo {author} {\bibfnamefont
  {Q.}~\bibnamefont {Liu}}, \bibinfo {author} {\bibfnamefont {Y.}~\bibnamefont
  {Shen}},\ and\ \bibinfo {author} {\bibfnamefont {X.}~\bibnamefont {Fu}},\
  }\bibfield  {title} {\enquote {\bibinfo {title} {Structured light analogy of
  quantum squeezed states},}\ }\href@noop {} {\bibfield  {journal} {\bibinfo
  {journal} {Light: Sci. Appl.}\ }\textbf {\bibinfo {volume} {13}},\ \bibinfo
  {pages} {297} (\bibinfo {year} {2024})}\BibitemShut {NoStop}%
\bibitem [{\citenamefont {Glorieux}\ \emph {et~al.}(2023)\citenamefont
  {Glorieux}, \citenamefont {Aladjidi}, \citenamefont {Lett},\ and\
  \citenamefont {Kaiser}}]{Glorieux_2023}%
  \BibitemOpen
  \bibfield  {author} {\bibinfo {author} {\bibfnamefont {Q.}~\bibnamefont
  {Glorieux}}, \bibinfo {author} {\bibfnamefont {T.}~\bibnamefont {Aladjidi}},
  \bibinfo {author} {\bibfnamefont {P.~D.}\ \bibnamefont {Lett}},\ and\
  \bibinfo {author} {\bibfnamefont {R.}~\bibnamefont {Kaiser}},\ }\bibfield
  {title} {\enquote {\bibinfo {title} {Hot atomic vapors for nonlinear and
  quantum optics},}\ }\href {https://doi.org/10.1088/1367-2630/acce5a}
  {\bibfield  {journal} {\bibinfo  {journal} {New J. Phys.}\ }\textbf {\bibinfo
  {volume} {25}},\ \bibinfo {pages} {051201} (\bibinfo {year}
  {2023})}\BibitemShut {NoStop}%
\bibitem [{\citenamefont {Bello}\ \emph {et~al.}(2021)\citenamefont {Bello},
  \citenamefont {Michael}, \citenamefont {Rosenbluh}, \citenamefont {Cohen},\
  and\ \citenamefont {Pe'er}}]{Bello.2021}%
  \BibitemOpen
  \bibfield  {author} {\bibinfo {author} {\bibfnamefont {L.}~\bibnamefont
  {Bello}}, \bibinfo {author} {\bibfnamefont {Y.}~\bibnamefont {Michael}},
  \bibinfo {author} {\bibfnamefont {M.}~\bibnamefont {Rosenbluh}}, \bibinfo
  {author} {\bibfnamefont {E.}~\bibnamefont {Cohen}},\ and\ \bibinfo {author}
  {\bibfnamefont {A.}~\bibnamefont {Pe'er}},\ }\bibfield  {title} {\enquote
  {\bibinfo {title} {Broadband complex two-mode quadratures for quantum
  optics},}\ }\href {https://doi.org/10.1364/OE.432054} {\bibfield  {journal}
  {\bibinfo  {journal} {Opt. Express}\ }\textbf {\bibinfo {volume} {29}},\
  \bibinfo {pages} {41282--41302} (\bibinfo {year} {2021})}\BibitemShut
  {NoStop}%
\bibitem [{Note1()}]{Note1}%
  \BibitemOpen
  \bibinfo {note} {In the fermionic ($SU(2)$) case this amounts to use the
  Euler parametrization $(\varphi ,\theta ,\psi )$ and keep only the rotation
  around $y$ by the angle $\theta $.}\BibitemShut {Stop}%
\bibitem [{\citenamefont {Van-Brunt}\ and\ \citenamefont
  {Visser}(2015)}]{van-brunt.visser.JPA.2015}%
  \BibitemOpen
  \bibfield  {author} {\bibinfo {author} {\bibfnamefont {A.}~\bibnamefont
  {Van-Brunt}}\ and\ \bibinfo {author} {\bibfnamefont {M.}~\bibnamefont
  {Visser}},\ }\bibfield  {title} {\enquote {\bibinfo {title} {{Special-case
  closed form of the Baker–Campbell–Hausdorff formula}},}\ }\href@noop {}
  {\bibfield  {journal} {\bibinfo  {journal} {J. Phys. A: Math. Theor.}\
  }\textbf {\bibinfo {volume} {48}},\ \bibinfo {pages} {225207} (\bibinfo
  {year} {2015})}\BibitemShut {NoStop}%
\end{thebibliography}%

\end{document}